\documentclass[a4paper]{aa}
\usepackage{graphics,txfonts,aalongtable}

\newcommand{\te}{$T_{\rm eff}$}
\newcommand{\tev}[1]{$T_{\rm eff}= #1$~K}

\newcommand{\dfrac}[2]{\frac{\displaystyle #1}{\displaystyle #2}}
\newcommand{\Bz}{\ensuremath{\langle B_{\rm z}\rangle}}
\newcommand{\bBz}{\ensuremath{\langle \overline{B_{\rm z}}\rangle}}
\newcommand{\cz}{\ensuremath{C_{\rm z}}}

\newcommand{\fifps}[2]{\centering\resizebox{#1}{!}{\includegraphics{#2}}}
\newcommand{\figps}[1]{\resizebox{\hsize}{!}{\rotatebox{0}{\includegraphics{#1}}}}

\newcommand{\beq}{\begin{equation}}
\newcommand{\eeq}{\end{equation}}

\begin{document}

\title{Evolutionary state of magnetic chemically peculiar stars
\thanks{Tables 1 and 2 are only available in electronic form at {\tt http://www.edpsciences.org}}\fnmsep
\thanks{Based on observations made with ESO Telescopes at the Paranal
        Observatory under programs ID 71.D-0308, 72.D-0377, and
        73.D-0464, retrieved through the ESO archive.}}

\author{O. Kochukhov\inst{1} \and S. Bagnulo\inst{2}}

\offprints{O. Kochukhov, \email{oleg@astro.uu.se}}

\institute{Department of Astronomy and Space Physics, Uppsala University, SE-751 20, Uppsala, Sweden
           \and European Southern Observatory, Casilla 19001, Santiago 19, Chile}

\date{Received 28 November 2005 / Accepted 16 January 2006}

\abstract 
{The photospheres of about 5--10\,\% of the upper main sequence stars
exhibit remarkable chemical anomalies. Many of these
chemically peculiar (CP) stars have a global magnetic field, the origin of
which is still a matter of debate.}
{We present a comprehensive statistical investigation of the evolution
of magnetic CP stars, aimed at providing constraints to the
theories that deal with the origin of the magnetic field in these
stars.}
{We have collected from the literature data for 150 magnetic CP stars
with accurate Hipparcos parallaxes. We have retrieved from the ESO
archive 142 FORS1 observations of circularly polarized spectra for 100
stars. From these spectra we have measured the mean longitudinal
magnetic field, and discovered 48 new magnetic CP stars (five of which
belonging to the rare class of rapidly oscillating Ap stars). We have
determined effective temperature and luminosity, then mass and
position in the H-R diagram for a final sample of 194 magnetic CP
stars.}
{We found that magnetic stars with $M > 3 \,M_\odot$ are
homogeneously distributed along the main sequence. Instead, there are
statistical indications that lower mass stars (especially those with
$M \le 2\,M_\odot$) tend to concentrate in the centre of the main
sequence band.  We show that this inhomogeneous age distribution
cannot be attributed to the effects of random errors and small number
statistics. Our data suggest also that the surface magnetic flux of CP
stars increases with stellar age and mass, and correlates with the
rotation period. For stars with $M > 3\,M_\odot$, rotation periods
decrease with age in a way consistent with the conservation of the
angular momentum, while for less massive magnetic CP stars an
angular momentum loss cannot be ruled out.}
{The mechanism that originates and sustains the magnetic field in the
upper main sequence stars may be different in CP stars of different mass.}


\maketitle

\section{Introduction}
\label{intro}
Observations suggest that magnetic fields are ubiquitous in late-type
stars, and that a correlation exists between magnetic activity and
stellar rotation. The magnetic field of late-type stars is typically
localised in spots, and evolves on relatively short time scales.
Although not yet fully understood, a \textit{dynamo} mechanism is
commonly invoked to explain the presence of a magnetic field in these
kinds of stars.

Early-type stars show a completely different magnetic phenomenology.
Magnetic fields appear organised on a large-scale at the stellar
surface, and do not change within a time scale shorter than
several decades. Instead, a periodic field variability is observed,
which is commonly interpreted in terms of the so-called \textit{Oblique
Rotator Model}: the magnetic field geometry is not symmetric about the
rotation axis, and the observer sees a magnetic configuration that
changes as the star rotates. The field strength (typically a few
hundreds up to a few tens of thousands of Gauss) does not seem
correlated to the star's rotational velocity. Only a minority
(about 5\,\%) of early-type stars is magnetic. Practically all 
known magnetic stars of the upper main sequence are classified
between late F- and early B-type, and belong to the category
of the so-called \textit{chemically peculiar} (CP) stars, i.e., stars
that exhibit distinctive peculiarities in the element abundance of their
atmospheres. Most CP stars (hence most magnetic stars) show also abnormally slow rotation.

The origin of the magnetic fields in CP stars is a matter of debate (Moss
\cite{DM04}). The dynamo hypothesis can hardly explain the observed high
field strengths and the lack of a correlation with rotation. A more
promising approach is offered by the \textit{fossil field} theory (Cowling
\cite{C45}; Moss \cite{DM89}; Braithwaite \& Spruit \cite{BS04}), according
to which the observed fields in the upper main sequence magnetic stars are
the remnants of fields present during earlier stages of stellar evolution.
The fact that no correlation is observed with stellar rotation, and the fact
that only a small percentage of the upper main sequence stars are magnetic,
are naturally explained in terms of variations in the amount of magnetic
flux trapped during star formation. However, it not yet clear
(observationally nor theoretically) if and how these fields evolve during
the main sequence phase.

To provide constraints to the theory of the origin of the magnetic field,
it is important to study the \textit{evolutionary state} of magnetic
CP stars. This has been done by several authors, and with conflicting
conclusions. Based on a small sample of about 30 magnetic stars, Hubrig et al.
(\cite{HNM00}) suggested
that magnetic fields appear at the surface of CP stars with $M\la 3\,M_\odot$
only after they have spent considerable fraction of their life on the main 
sequence. Most low mass stars studied by Hubrig et al. (\cite{HNM00}) were drawn from the
group of very slowly rotating magnetic stars with detectable Zeeman
splitting in their spectra (Mathys et al. \cite{MHL97}). This selection procedure 
results in a statistic sample small in size, that may also be intrinsically 
biased toward older stars (should the star's angular momentum be lost or/and
the surface magnetic field increase
during the main sequence evolutionary phase). Both Gomez et al.~(\cite{GLG98}) and
P\"{o}hnl et al.~(\cite{PPM05}) investigated much larger samples of CP stars,
but instead of selecting objects for which magnetic field was detected, 
they utilised chemical peculiarity as a proxy of magnetism. They have found that
CP stars with chemical anomalies similar to those observed in magnetic CP
stars occupy all the regions of the main sequence. 

In an attempt to clarify this puzzling situation, we have carried out a new
study of the evolution of magnetic stars of the upper main sequence that is
based on a very large sample of observed magnetic CP stars, collected
through a thoroughfull investigation of the literature and of the ESO
archive. Our study includes all known objects for which precise parallaxes
were measured by the Hipparcos mission, and for which the presence of 
magnetic field at the surface could be asserted via direct field detections.
We have also investigated whether there are observational evidences for
angular momentum losses occurring during the star's evolution in the main
sequence.

This paper is organised as follows. Section~\ref{sample} describes the
observations collected from the literature, and presents new magnetic
measurements obtained from the analysis of data collected with FORS1 at
the ESO VLT. In Sect.~\ref{teff-lum} we determine the temperature
and luminosity of the selected stellar sample, and in Sect.~\ref{tracks}
we derive the position of the selected stars in the H-R diagram. Results
and statistical analysis are presented in Sect.~\ref{results} and
discussed in Sect.~\ref{discuss}.

\section{The stellar sample}
\label{sample}

\subsection{Known magnetic stars}

The main part of our sample of magnetic CP stars has been drawn from
various literature sources.  The catalogues by Romanyuk (\cite{R00})
and Bychkov et al. (\cite{BBM03}) have provided the core of the sample
of the early-type stars with definite magnetic field detections. Since
the present study focuses on the classical magnetic Ap/Bp stars, we
have selected objects with the SrCrEu, Si, and/or He chemical
peculiarities for which reliable detection of magnetic field has been
obtained with one of the two direct methods: spectropolarimetric or
photopolarimetric measurement of the disk-averaged line of sight
(longitudinal) magnetic field (Mathys \cite{M91}; Borra \& Landstreet
\cite{BL80}) and the mean field modulus diagnostic using resolved
Zeeman split spectral line profiles (Mathys et al. \cite{MHL97}).

The list of stars compiled from the two catalogues was
supplemented by a number of recently discovered magnetic stars. In
particular, the studies by Auri\`{e}re et al. (\cite{ASW04}), El'kin et al.
(\cite{EKR03}), Johnson (\cite{J04}), Hubrig et al. (\cite{HSS03,HNS05}), 
Ryabchikova et al. (\cite{RLK04,RWA05}), Shorlin et al. (\cite{SWD02}), and
St\"utz et al. (\cite{SRW03}) have provided objects satisfying our selection
criteria. We refer the reader to the original papers for the description
of the respective observations and their analysis. Putting aside details
of these studies of individual magnetic CP stars, it is worth noting that
all investigations mentioned above relied on the classical direct methods
of the magnetic field detection (spectropolarimetry and/or measurement of
the resolved Zeeman split lines) and that only objects with reliable
($>$\,3$\sigma$ in the case of longitudinal field measurements) magnetic
field detections were included in our sample. 

According to the current understanding of the incidence of magnetism
in early-type stars, the non-magnetic chemically peculiar stars,
including objects with the HgMn and Am peculiarity types, do not host
global magnetic fields similar to those found in the classical
magnetic CP stars. Although claims of the detection of
magnetic field in HgMn or Am stars have occasionally been made (e.g.,
Mathys \& Lanz \cite{ML90}; Hubrig \& Castelli \cite{HC01}), these
studies typically employed an indirect diagnostic of magnetic field
based on the analysis of magnetic broadening and intensification of
spectral lines. Furthermore, none of the alleged detections of the
complex fields in non-magnetic CP stars was confirmed by a subsequent,
independent observation or supported by direct spectropolarimetric
magnetic field diagnostic technique (Shorlin et al. \cite{SWD02}; Wade
et al. \cite{WAD03}). Given this absence of the solid evidence for the
presence of significant surface fields in HgMn and Am stars,
all such objects were eliminated from our list of magnetic
stars. We emphasise that in nearly all cases when non-zero
longitudinal field determinations are reported for HgMn or Am stars in
the catalogues compiled by Romanyuk (\cite{R00}) and Bychkov et
al. (\cite{BBM03}), the original measurements date back to the early
low-precision spectropolarimetric observations by Babcock
(\cite{B58}).

The magnetic stellar sample compiled in the present study was cross-matched
with the Hipparcos parallax data (Perryman et al. \cite{PLK97}). Parallaxes
with an accuracy better than 20\% (which is the same threshold as employed
by Hubrig et al. \cite{HNM00}) could be retrieved for 150 magnetic CP
stars. Magnetic field in the majority of these stars was detected using
polarimetric observations. Consequently, unlike the 33 magnetic stars
studied by Hubrig et al. (\cite{HNM00}), our sample does not suffer from 
the possible bias of over-representing the slowly rotating and strongly magnetic 
old stars for which resolved Zeeman split lines are observable.

\subsection{Magnetic stars observed with FORS1 at VLT}
\label{fors}

FORS1 (FOcal Reducer/low dispersion Spectrograph) of the ESO VLT is a
multi-mode instrument equipped with polarisation analysing optics
including super-achromatic half-wave and quarter-wave phase retarder
plates, and a Wollaston prism.
In the last few years, FORS1 has been extensively used in polarimetric
mode to measure magnetic fields in various kinds of stars, for instance
in white dwarfs (e.g., Aznar Cuadrado et al. \cite{AJN04}), in hot
subdwarfs (O'Toole et al. \cite{OJF05}), in the central stars of
planetary nebulae (Jordan et al. \cite{JWO05}). In particular,
several observing programs 
have been dedicated to the magnetic field
surveys of Ap stars, and many data are now available in the ESO
archive. For this work we have retrieved 142 observations of circular
polarised spectra for a total of 100 Ap stars. 
All targets have been observed using grism 600\,B and a $0.4''$
slit-width, which gives a spectral resolution of about 2000, covering
the spectral range 3550--5850\,\AA.  A typical observing block
consists of four frames obtained with the $\lambda/4$ retarder
waveplate oriented at $+45\degr$ and four frames obtained with the
retarder waveplate at $-45\degr$. Data have been reduced using standard IRAF routines
in order to get wavelength calibrated spectra from the ordinary and
extra-ordinary beams taken with the $\lambda/4$ retarder waveplate
oriented at $+45\degr$ and $-45\degr$.  From these spectra we have
obtained Stokes $I$ and $V$ profiles using Eq.~(4.1) of the FORS1+2
User Manual (VLT-MAN-ESO-13100-1543), but with the sign changed. We
have then considered the Stokes $I$ and $V$ profiles of the Balmer
lines from H$\beta$ down to the Balmer jump, and then we have obtained
the longitudinal field \Bz\ using a least-square technique based on
the formula
\begin{equation}
\frac{V}{I} = - g_\mathrm{eff} \ \cz \ \lambda^{2} \
                 \frac{1}{I} \
                 \frac{\mathrm{d}I}{\mathrm{d}\lambda} \
                 \Bz\;,
\label{EqBz}
\end{equation}
where $g_\mathrm{eff}$ is the effective Land\'{e} factor ($=1$ for the hydrogen
Balmer lines, see Casini \& Landi Degl'Innocenti \cite{CL94}),
$\lambda$ is the wavelength expressed in \AA, \Bz\ is the longitudinal
field expressed in Gauss, and
\[
\cz = \frac{e}{4 \pi m_\mathrm{e} c^2}
\ \ \ \ \ (\simeq 4.67 \times 10^{-13}\,\mathrm{\AA}^{-1}\ {\rm G}^{-1}),
\]
where $e$ is the electron charge, $m_\mathrm{e}$ the electron mass,
$c$ the speed of light. Equation~(\ref{EqBz}) is valid under the
weak-field approximation, which, for the hydrogen Balmer lines formed in a typical
atmosphere of an A-type star, holds for field strength up to $\sim
20$\,kG. More details on the data reduction technique, and the way
\Bz\ is calculated, are given in Bagnulo et al. (\cite{BSW02}) and
Bagnulo et al. (\cite{BLM05}).

The new \Bz\ measurements are reported in Table~\ref{tbl1} 
(available in electronic form only).  Magnetic
field was detected at $>3\sigma$ level in 53 stars. Only five of them
were previously known to be magnetic, whereas the other four lack
accurate parallax data in the Hipparcos catalogue. Thus, analysis of
the archival FORS1 spectra has contributed with 44 magnetic stars, increasing
the total sample investigated in the present paper to 194 objects.

We note that definite detection of the longitudinal field in five members
(HD\,19918, HD\,42659, HD\,60435, HD\,84041, HD\,86181) of the
rare class of rapidly oscillating Ap (roAp) stars is reported here for the
first time, thereby significantly increasing the number of roAp stars with
known magnetic field properties.

\section{Determination of effective temperatures and luminosities}
\label{teff-lum}

\subsection{Photometric effective temperature}
\label{teffder}

Stellar effective temperatures were determined using calibration of the
Geneva photometric system (Golay \cite{G72}). Observed photometric
parameters were extracted from the catalogue of Rufener (\cite{R89}) and
supplemented by the data available through the online photometric database
at Geneva  Observatory\footnote{{\tt
http://obswww.unige.ch/gcpd/ph13.html}}. We determined \te\ of CP stars
following the procedure suggested by Hauck \& North (\cite{HN93}) and
revised by Hauck \& K\"unzli (\cite{HK96}). For hot stars calibration in the
theoretical grids published by K\"unzli et al. (\cite{KNK97}) was used in
combination with the linear \te\ correction to account for anomalous flux
distribution of magnetic CP stars (Hauck \& K\"unzli \cite{HK96}). For cool
Ap stars we employed calibration of the $(B2-G)_0$ color index proposed by
Hauck \& North (\cite{HN93}).

For a few CP stars lacking photometric measurements in the Geneva system we
determined \te\ using Str\"omgren $uvby\beta$ photometric data (Hauck \&
Mermilliod \cite{HM98}) and calibration by Moon \& Dworetsky (\cite{MD85}).

Effective temperature of the extreme cool magnetic peculiar star HD\,101065
(Przybylski's star) cannot be determined using any usual calibrations
available either for normal or CP stars. Instead, a \tev{6450} was adopted
for this object based on the results of recent detailed spectroscopic
studies (Cowley et al. \cite{CRK00}; Kochukhov et al. \cite{KBB02}).
The spectroscopic \tev{7750} (Kochukhov et al. \cite{KBB02}) was also 
used for HD\,216018, which lacks a complete set of Str\"omgren or Geneva photometry.

For a subsample of stars with both Str\"omgren and Geneva photometry
available, uncertainty of effective temperature can be estimated from the
discrepancy of \te\ values given by independent calibrations in the two
different photometric systems. Based on this assessment we adopted
$\sigma(T_{\rm eff})=200$~K for  stars with $T_{\rm eff}\le8500$~K,
$\sigma(T_{\rm eff})=300$~K for $8500<T_{\rm eff}\le10500$~K,
$\sigma(T_{\rm eff})=400$~K for $10500<T_{\rm eff}\le16000$~K, and
$\sigma(T_{\rm eff})=500$~K for stars hotter than \tev{16000}. These error
estimates are very similar to the \te\ uncertainty assumed by Hubrig et al.
(\cite{HNM00}, see their tables 1 and 2), although no explicit discussion of
the adopted \te\ error bars can be found in the latter paper. 

\subsection{Correction for the interstellar extinction and reddening}

The interstellar extinction and reddening have to be taken into account for
stars located farther away than  60~pc. We considered four different
procedures to obtain color excess $E(B-V)$ for individual stars in our
sample: from the intrinsic $[U-B]$ color and reddening-free Geneva $X$ and
$Y$ parameters of hotter stars  (Cramer \cite{C82}), from the interstellar
extinction maps of Lucke (\cite{L78}) and Schlegel et al. (\cite{SFD98}),
and using the model of Hakkila et al. (\cite{HMS97}). The $E(B-V)$
parameters obtained from these sources were averaged after rejection of
occasional outliers. Based on the scatter of the color excess values derived
using different methods, we found a typical uncertainty of $E(B-V)$ to be
0.005~mag for $E(B-V)\le0.05$ and 0.010~mag for $E(B-V)>0.05$. For several
strongly reddened objects a higher $E(B-V)$ error bar had to be adopted,
reflecting large standard deviation of highly discrepant reddening
estimates.

Photometric parameters in the Geneva and Str\"omgren system were dereddened
with the help of relations based on the interstellar extinction laws given
by Fitzpatrick (\cite{F99}). Interstellar extinction in the $V$-band was 
calculated using $R\equiv A_V/E(B-V)=3.1$.

\subsection{Hipparcos luminosity}

\begin{figure}[!th]
\figps{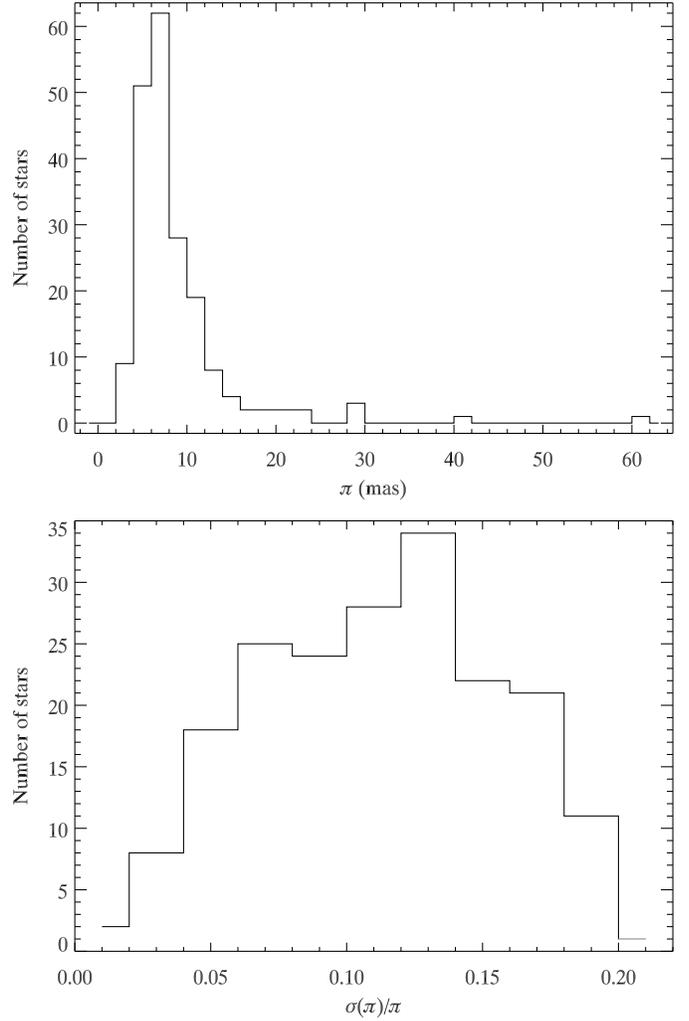}
\caption{Distribution of the Hipparcos parallaxes (\textit{top}) and their
relative errors (\textit{bottom}) for the sample of magnetic CP stars.}
\label{fig1}
\end{figure}

Absolute magnitudes and luminosities of the program stars were determined on
the basis of data from the Hipparcos catalogue (Perryman et al. \cite{PLK97}) and the
$V$-magnitude information extracted from the {\sc simbad} database.
Distribution of the Hipparcos parallaxes and their relative errors is
illustrated in Fig.\,\ref{fig1}. Most of the studied stars are located
within 250 pc from the Sun, but only a few are closer than 60 pc. The
average parallax uncertainty for the magnetic stars in our sample is 11\%.
Roughly half of the stars have parallax determined  with an accuracy better
than that. 

\begin{figure*}[!t]
\fifps{15cm}{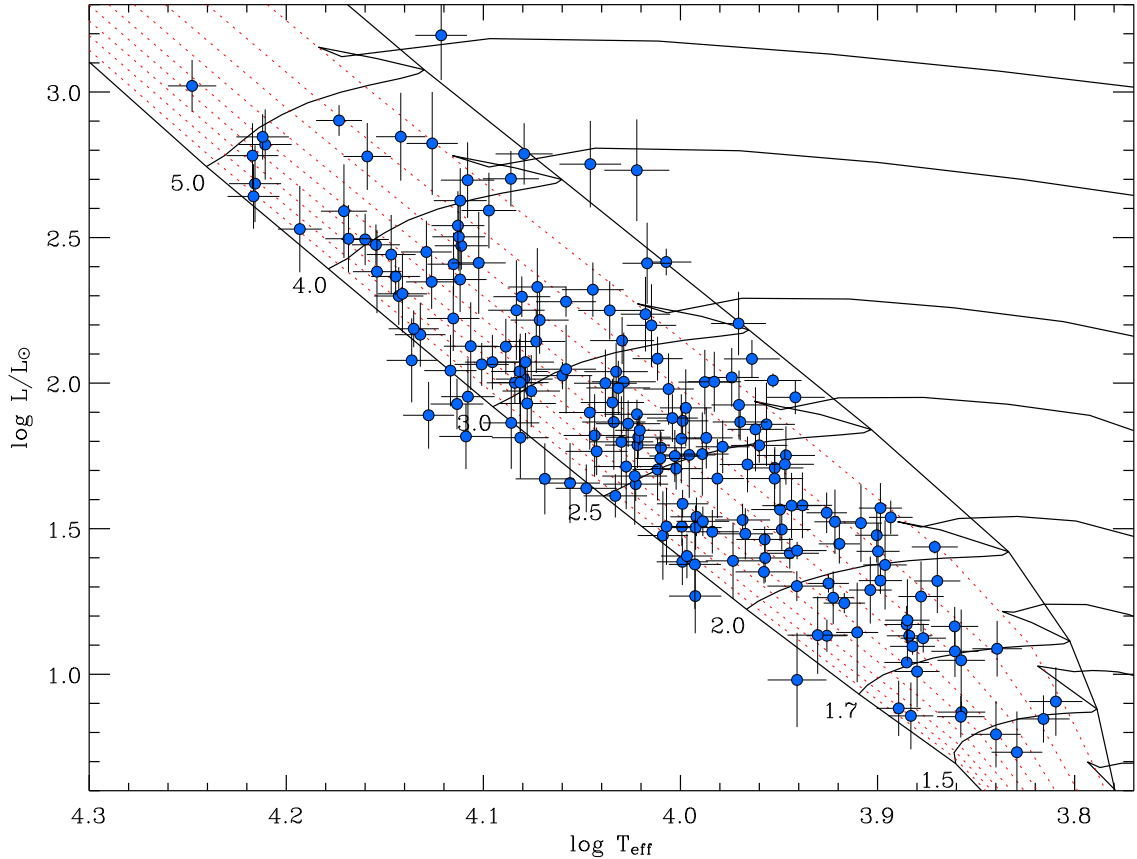}
\caption{
Position of magnetic chemically peculiar stars on the Hertzsprung-Russell
diagram. The solid lines show theoretical evolutionary tracks, the ZAMS and the
envelope of lowest \te\ achieved during the main sequence evolutionary stage (Schaller et
al. \cite{SSM92}; Schaerer et al. \cite{SCM93}). Dotted curves correspond to
the lines of equal fractional age $\tau$ measured in units of the main sequence
stellar lifetime. These curves are plotted with a step of 0.1 between the ZAMS
($\tau = 0$) and the end of the core hydrogen burning phase ($\tau = 1$).}
\label{fig2}
\end{figure*}

\begin{figure}[!t]
\figps{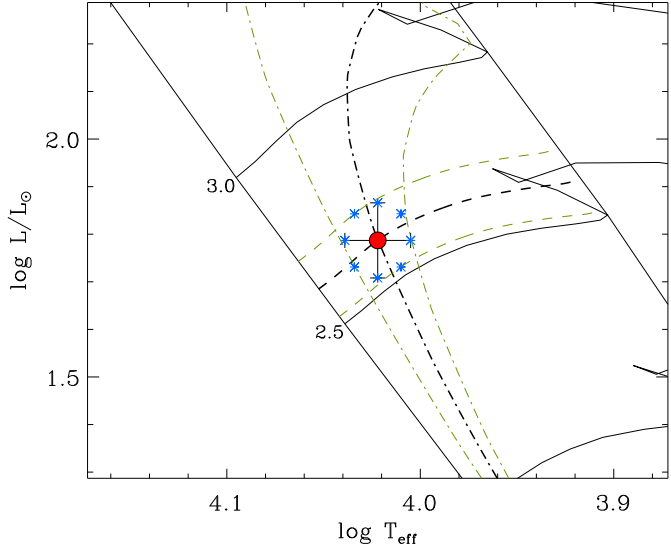}
\caption{Illustration of the error analysis applied to derive asymmetric
confidence intervals for the stellar mass and age. The solid lines show
theoretical evolutionary tracks  (Schaller et al. \cite{SSM92}; Schaerer et
al. \cite{SCM93}). The point and error bars indicate our estimate of the
temperature and luminosity of the magnetic CP star HD\,56350. Thick dashed and
dash-dotted lines correspond to interpolated evolutionary tracks and
isochrones,  respectively. Interpolation procedure is repeated for 8
locations along the error ellipse in the  $\log\,T_{\rm
eff}$--$\log\,L/L_{\sun}$ plane, as shown by asterisks. Thin dashed and
dash-dotted curves show tracks and isochrones for minimum and maximum mass
and age. This analysis yields a stellar mass 
$M/M_{\sun}$\,=\,$2.610^{+0.090}_{-0.088}$ and age
$\log\,t$\,=\,$8.40^{+0.11}_{-0.21}$ yr.}
\label{fig3}
\end{figure}

\begin{figure}[!t]
\figps{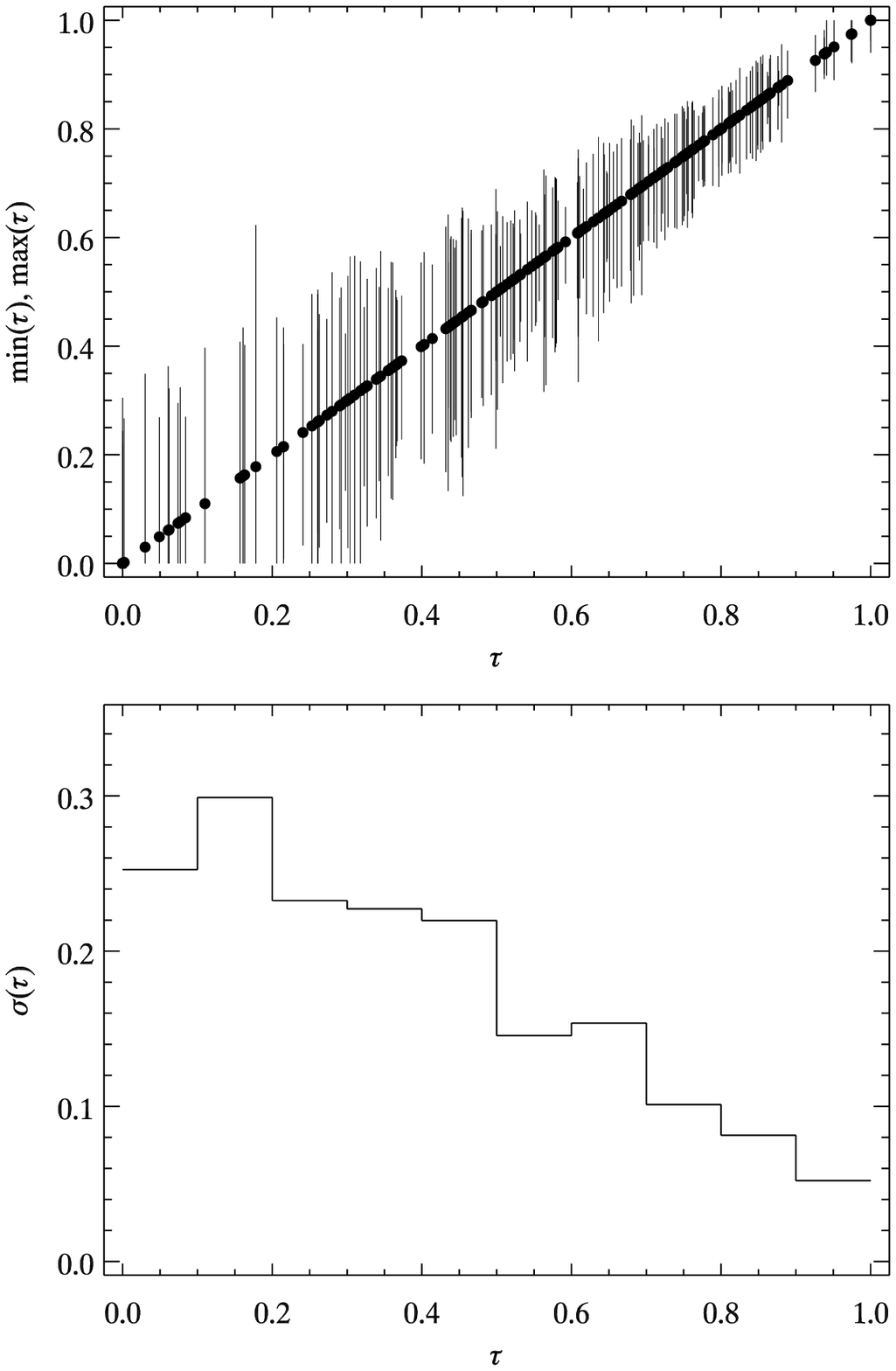}
\caption{Distribution of 1$\sigma$ confidence intervals of the fractional ages
(\textit{top}) and respective mean error bar  (\textit{bottom}) as a
function of the relative stellar age for the studied sample of magnetic stars.}
\label{fig4}
\end{figure}

From the comparison of magnitudes given in different literature sources, we
estimated uncertainty of $m_V$ related to the measurement errors and
intrinsic stellar variability to be approximately 0.02~mag. 

For a number of magnetic stars known to be members of spectroscopic or
unresolved visual binaries we applied positive duplicity correction to
$m_V$. Generally, this correction can be estimated from the luminosity
ratios available from literature for the majority of well-studied SB2 and
visual binaries. On the other hand, CP stars belonging to SB1 systems usually lack precise
luminosity ratio estimates. For  these objects we adopted $\Delta m_V=0.16$
which corresponds to the magnitude difference of 2.0 between components. 

The absolute magnitude in the $V$-band was determined using the standard
relation:
\beq
M_V = m_V + 5 + 5\/\log\pi - A_V,
\eeq
where trigonometric parallax $\pi$ is measured in arcseconds and
interstellar extinction was determined as outlined above. The error estimate
for the absolute magnitude took into account uncertainties in $m_V$, $\pi$,
and $A_V$:
\beq
\sigma(M_V) = \sqrt{\sigma^2(m_V) + \left(\dfrac{5\/\sigma(\pi)}{\pi \ln10}\right)^2 + \sigma^2(A_V)}.
\eeq

Calculating the stellar luminosity,
\beq
\log\dfrac{L}{L_{\sun}} = - \dfrac{M_V + BC - M_{\rm bol}(\sun)}{2.5},
\eeq
we adopted the solar bolometric magnitude $M_{\rm bol}(\sun)=+4.75$ (Bessell
\cite{B00}) and used the standard bolometric correction $BC$ taken from
Flower (\cite{F96}). Since in the latter paper $BC$ is tabulated as a
function of \te, uncertainty of effective temperature contributes to the
total error of luminosity that can be estimated according to the following
expression:
\beq
\sigma\left(\log\dfrac{L}{L_{\sun}}\right) = 
0.4\/\sqrt{\sigma^2(M_V) + \left(\dfrac{\mathrm{d}BC}{\mathrm{d}T_{\rm eff}}\right)^2 \sigma^2(T_{\rm eff})}.
\eeq
Taking into account all contributions to the $M_V$ and $L/L_{\sun}$ error
budgets, we find a typical  uncertainty of 20--25\% for both parameters.

Anomalous flux distribution of peculiar stars is characterized by the
enhanced ultraviolet absorption which induces backwarming in the visible
(Leckrone \cite{L73}). This makes CP stars sub-luminous for their visual
colours, and bolometric correction has to be modified accordingly (North
\cite{N81}; Lanz \cite{L84}). In the present paper we use $BC$ tabulated as
a function of \te\ for normal stars (Flower \cite{F96}). However, we
account for peculiar nature of CP stars in derivation of their  \te\
(Sect.~\ref{teffder}). With this procedure the average change in $BC$ due to
anomalous stellar flux distribution is taken into account implicitly.
Remaining modifications of $BC$ depend on individual properties of magnetic
stars and are  difficult to estimate without detailed model
atmosphere analysis. Nevertheless, this $BC$ uncertainty is likely to be
much smaller compared to the average error of $M_{\rm bol}$ (0.24~mag) for
the stars in our sample.

No Lutz-Kelker (LK, Lutz \& Kelker \cite{LK73}) correction was applied to
the absolute magnitudes of magnetic stars in our sample. The original LK
correction, which is always negative and hence systematically increases
luminosity estimated from parallax, has been a source of confusion and
extensive debate in literature. In a recent publication Smith (\cite{S03})
summarised this discussion and concluded that the LK correction is meaningful
only when applied to the stellar samples, but should not be used in studies
of individual stars. As pointed out by St\c{e}pie\'{n} (\cite{S04}), large
(from $-0.1$~mag up to $-0.5$~mag) negative LK corrections adopted by
Hubrig et al. (\cite{HNM00}) displaced the stars in their sample
significantly upwards from the zero age main sequence (ZAMS), 
possibly resulting in apparent lack of young
low mass magnetic stars. A similar problem may have affected results of
P\"ohnl et al. (\cite{PPM05}) who, contrary to the recommendations of Smith
(\cite{S03}) and  St\c{e}pie\'{n} (\cite{S04}), have also attempted to use the
LK procedure to correct individual stellar magnitudes.

\section{Mass and age determination}
\label{tracks}

Theoretical evolutionary tracks of the upper main sequence stars are
available for different metal abundance of the stellar envelope (see
Schaller et al. \cite{SSM92} and Schaerer et al.  \cite{SCM93}). Large
deviations of the chemical composition of CP stars from the normal solar
abundance table are believed to be limited to the surface layers. These
superficial chemical anomalies are produced  by the process of selective
radiative diffusion in the presence of magnetic field, stellar wind, and
possibly weak turbulent mixing. These complex hydrodynamical effects are
poorly understood and hence it is not possible to estimate the average
interior metal content of individual CP stars based on the observed surface
abundance pattern. In this situation a fixed, usually solar, metallicity has
to be assumed to make comparison of the observed and predicted stellar
parameters possible (e.g., Hubrig et al. \cite{HNM00}; P\"ohnl et al.
\cite{PPM05}). Here we have adopted metallicity $Z=0.018$ and obtained
theoretical stellar evolutionary tracks by interpolating within the grids of
Schaerer et al. (\cite{SCM93}) and Schaller et al. (\cite{SSM92}), who
published calculations for $Z=0.008$ and $Z=0.020$, respectively. The
plausible effect of the dispersion in $Z$ can be estimated from the scatter in
the surface abundances determined for normal B stars and nearby young F and
G stars. Using the summary of the Fe and light element abundances given by
Sofia \& Meyer (\cite{SM01}), we obtain star-to-star $Z$ variation of
$\approx0.002$. The resulting effect on the age and mass determination of 
CP stars is smaller than other error sources.

Figure~\ref{fig2} illustrates distribution of magnetic CP stars in the
theoretical H-R diagram. Each star is shown with a point, whereas
respective error bars give uncertainty of \te\ and luminosity. The
ZAMS and the envelope of the lowest \te\ achieved during the core
hydrogen burning phase are also shown. Given a set of theoretical
isochrones, the problem of the stellar mass determination reduces to
interpolation within the evolutionary tracks tabulated for different masses. To
avoid degeneracy, all stars were assumed to lie within the main sequence
zone where \te\ decreases monotonously. For a few objects located below the ZAMS
or above the lowest \te\ line, parameters were calculated for the closest
main sequence point.

Subsequent determination of the stellar age is complicated by the uneven
evolution of stars in the $\log\/T_{\rm eff}$--$\log\/L/L_{\sun}$ plane. Young
stars located close to the ZAMS change their temperature and luminosity very
slowly. The pace of evolution increases rapidly as the star ages and shifts
towards the terminal age main sequence. Therefore, the same uncertainty of
\te\ and $L$ translates into dramatically different age errors, depending on
whether the star is young  or evolved. Although this effect is well-known, to
our knowledge, no attempt has ever been made to take it properly into account
in the statistical studies of the evolutionary state of CP stars. In the present paper
we developed a non-linear error propagation procedure to obtain realistic
errors of the absolute and relative ages. For each star in our sample we
determined mass and age for the point corresponding to the adopted stellar
$\log\/T_{\rm eff}$ and $\log\/L/L_{\sun}$ and then repeated  this procedure
for 8 positions along the error ellipse defined by the individual uncertainty
of temperature and luminosity. Resulting minimum and maximum ages yield
realistic asymmetric range of evolutionary stages compatible with a given pair
of \te\ and $L$ and their respective 1$\sigma$ error limits. Application of
this non-linear error propagation procedure to the magnetic CP star HD\,56350
is illustrated in Fig.~\ref{fig3}. 

The summary of the age confidence limits is given in Fig.~\ref{fig4}. Here and
elsewhere in the paper we quantify the relative stellar age, $\tau$, by the
fraction of the stellar life spent on the main sequence. The ZAMS line corresponds
to $\tau=0$, whereas the star at the end of the core hydrogen burning phase
has $\tau=1$. As it follows from Fig.~\ref{fig4}, the relative age of individual young CP
stars cannot be determined with an accuracy better than $\approx$\,20\,\%.
Only for the oldest stars in our sample the H-R diagram fitting yields ages
precise at the 5--7\,\% level. 

The final set of the stellar fundamental parameters and corresponding 1$\sigma$
(68\,\%) confidence intervals is reported in Table~\ref{tbl2} (available in 
electronic form only). For each star
we list its identification in the HD and Hipparcos catalogues, distance,
absolute magnitude, effective temperature, luminosity, mass, absolute and
relative ages. Using the data in Table~\ref{tbl2}, is also
straightforward to estimate the stellar radius
\beq
\log(R/R_\odot) = 0.5 \log(L/L_\odot) - 2\log(T_{\rm eff}/T_{\rm eff\odot})
\eeq
and the surface gravitational acceleration
\beq
\log (g/g_\odot) = \log(M/M_\odot) + 4 \log(T_{\rm eff}/T_{\rm eff\odot})-\log(L/L_\odot).
\eeq

\begin{figure}[!t]
\figps{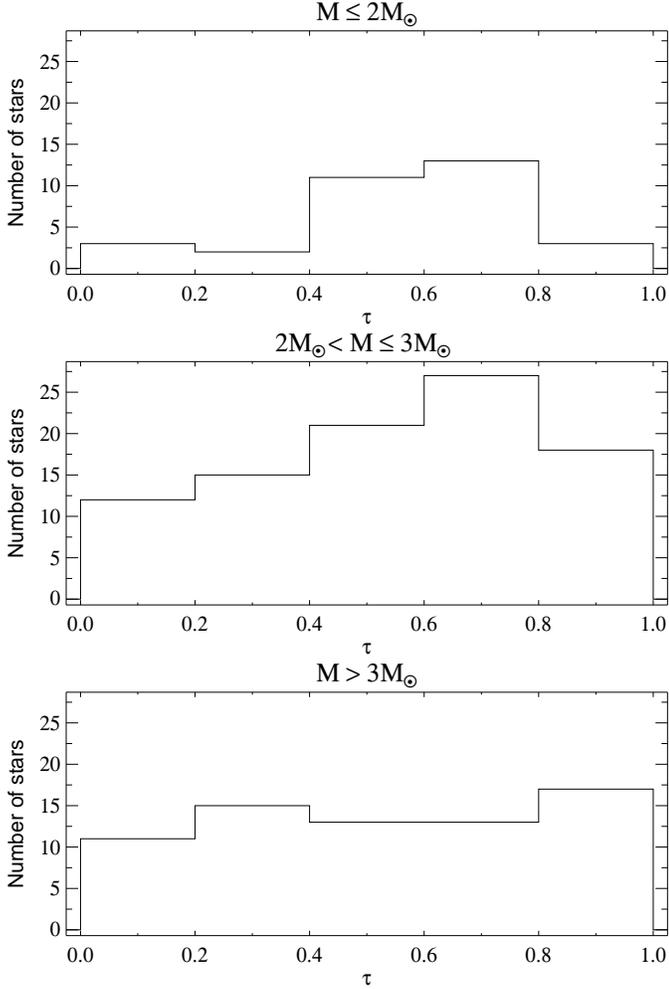}
\caption{Distribution of the relative ages for magnetic CP stars of 
different mass.}
\label{fig5}
\end{figure}

\section{Results}
\label{results}

\subsection{Distribution of stars in the H-R diagram}

\begin{figure}[!t]
\figps{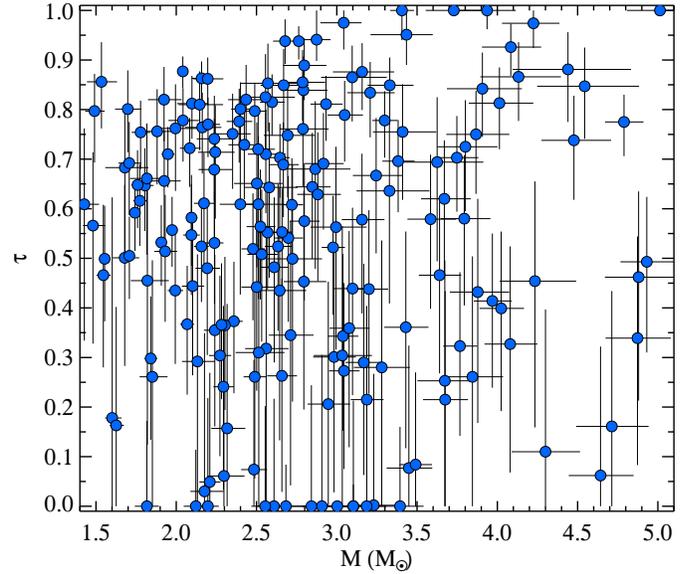}
\caption{Distribution of magnetic CP stars in the $\tau - M$ plane.}
\label{fig6}
\end{figure}

We have grouped all magnetic stars in our sample into three different
mass bins: stars with $M \le 2 M_\odot$, stars with $2 M_\odot < M \le
3 M_\odot$, and stars with $M > 3 M_\odot$.  Figure~\ref{fig5} shows
the distribution of the relative ages for the stars belonging to these
three groups. An alternative overview of the age distribution of the
magnetic stars of various masses is given in Fig.~\ref{fig6}.  We
found that 26\,\% of stars with mass $> 3\,M_\odot$ have spent less
than 30\,\% of their life in the main sequence. This percentage is
18\,\% for stars with mass between $2\,M_\odot$ and $3\,M_\odot$, and
only 16\,\% for stars with $M \le 2\,M_\odot$. It appears thus that
higher mass stars ($M>3 M_\odot$) are homogeneously distributed in
fractional age. In the group of stars of intermediate mass, younger
stars seem less numerous than older stars. Among stars with $M \le
2\,M_\odot$ the shortage of young objects is even more pronounced.

The statistical validity of these results must be carefully
investigated, since \textit{a priori} one could argue that a small
sample of objects randomly selected from a homogeneous group may well
be characterized by special features that in fact are not
representative of the entire sample. In other words, one could suspect
that the apparent shortage of young magnetic stars with $M \le
3\,M_\odot$ is just the result of a statistical fluctuation. It should
also be noted that young stars are those for which age determination
is less accurate. In order to clarify this situation, we have
performed numerical simulations to calculate what is the probability
that, repeating the same study, one obtains a number of stars with
$\tau < 0.3$ equal or smaller than what we have found. This 
estimates the false alarm probability, i.e. the chance that
the observed uneven distribution of stellar ages can be attributed 
to a statistical fluctuation.

First of all, we have given an analytical representation to the error
bars of the fractional age, by using a linear fit to the values given in
Table~\ref{tbl2}. Namely, we have estimated the lower and upper error
bars as
\begin{equation}\label{errorbars}
\Delta \tau_{-} = 0.33 - 0.29\, \tau \ \ \mathrm{and}\ \ \ 
\Delta \tau_{+} = 0.27 - 0.26\, \tau,
\end{equation}
respectively. Then we have considered a sample of $N$ objects. To each
of them, we have associated a random number $p_i$ from 0.0 to 1.0,
representing the fractional age. Each $p_i$ number has been
transformed to a $p_i' = p_i + \delta(p_i)$ value, where the
``errors'' $\delta(p_i)$ were deduced from a Gaussian distribution with
standard deviation $\Delta \tau_{-}$ (for $\delta(p)<0$) or $\Delta \tau_{+}$
(for $\delta(p)>0$), using again a random number generator. All $p_i' <
0$ have been then set to 0, and all $p_i' > 1$ have been changed to
1. We have then counted the number $J_k$ of $p_i'$ values included
within the interval $[0.0,0.3]$. We have repeated the same exercise $M$
times, and finally counted the number $L$ of times in which $J_k$ was
equal or smaller than a certain number $Q$. The ratio $P_1(N,Q) = L/M$ gives the
probability that, in a sample of N stars homogeneously distributed in
fractional age, we find no more than $Q$ stars in the interval $[0.0,0.3]$.

In our observational sample of $N=32$ magnetic stars with
$M\le2M_\odot$, we have found only 5 stars with $\tau \le
0.3$. Therefore we have performed the statistical test described above
using $N=32$ and $Q=5$, and calculated $P_1(32,5) = 6.5$\,\%.  We have
also found that in the sample of 93 magnetic stars with $2 M_\odot < M
\le 3\,M_\odot$, 17 have $\tau \le 0.3$. For this case, the
statistical test gives $P_1(93,17) = 1.3$\,\%. Finally, we have found
that 18 over 69 magnetic stars with $M>3\,M_\odot$ have $\tau \le
0.3$, and we have calculated $P_1(69,18) = 36$\,\%.

It is also of interest to test a complementary hypothesis that magnetic 
stars belong to a homogeneous population of objects with a relative age of at 
least $\tau=0.3$, and all young stars in our sample appear entirely due to
observational errors. In order to investigate this possibility, we have repeated
the simulations choosing the fractional age randomly in the interval $[0.3,1.0]$ and
counting  trials in which the number stars with $\tau<0.3$ has reached the
observed value. The resulting probability, $P_2(N,Q)$, turns out to be
considerable for stars with $M\le2M_\odot$ ($P_2=23$\,\%) but is negligible
($P_2\la 1$\,\%) for more massive stars.

Another conspicuous feature of the stellar age distributions presented in
Figs.~\ref{fig5} and \ref{fig6} is the relatively small number of
$M\le3\,M_\odot$ stars at the end of their main sequence life. Stellar
evolution is fast in this region of the H-R diagram. Consequently, the relative
age is determined with good precision (see Fig.~\ref{fig4}). Applying the same
statistical approach as outlined above (for stars in the $[0.0,1.0]$ age
interval) we could, however, verify that the lack of stars with $\tau\ge 0.8$
is not particularly significant by itself: the corresponding false alarm
probability, $P_3(N,Q)$ is 7\,\% and 40\,\% for the two groups of low mass stars
($M \le 2\,M_\odot$ and $2\,M_\odot < M \le 3\,M_\odot$, respectively), and
$P_3=81$\,\% for stars with $M>3\,M_\odot$. 

\begin{figure}[!t]
\figps{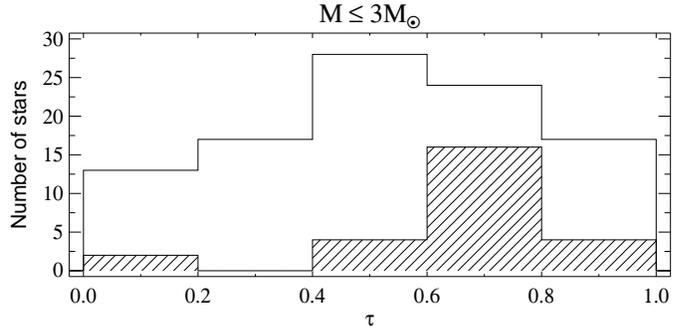}
\caption{Distribution of the relative ages for magnetic CP stars in the 
$M\le3\,M_\odot$ mass group. The hatched histogram shows the age 
distribution for stars with the resolved Zeeman split lines,
whereas the other histogram corresponds to stars without detectable 
Zeeman splitting.}
\label{fig7}
\end{figure}

We have further extended the statistical analysis to test the possibility 
that the \textit{overall} shape of the observed distribution of stellar ages
can be attributed to random errors and effects of small number statistics.
This was achieved by computing the composite probability that, given the 
number of objects observed in each mass range, the fraction of stars in  the
relative age interval $[0.0,0.4]$ (the sum of the first two bins in the
histograms of Fig.~\ref{fig5}) and, simultaneously, of those in the
$[0.8,1.0]$ interval  (the last bin in Fig.~\ref{fig5}) does not exceed the
observed values. We have found that the observed strong concentration of stars
in the middle of the H-R diagram is seldom realised in a random sample of
low mass stars drawn from a homogeneous age distribution.  Denoting the
corresponding probability with $P_4$, we have obtained $P_4=0.004$\,\% and
$P_4=0.6$\,\% for stars with $M \le 2 M_\odot$ and $2 M_\odot < M \le
3 M_\odot$, respectively. At the same time, for stars with $M>3\,M_\odot$ we
have determined $P_4=39$\,\%.

The conclusion of this series of tests is that it is very unlikely that 
the picture we have found is due to random errors in the determination of the 
stellar fundamental parameters. Unless the methods employed to measure stellar 
temperature and luminosity are affected by some large systematic errors, and 
assuming that the evolutionary models are correct, the following scenario
emerges from our study.
Magnetic stars with $M \le 3 M_\odot$ are concentrated in the centre
of the H-R diagram, and, in this mass range, older stars are more
numerous than younger magnetic stars. In particular, the age distribution of
stars with $M \le 2 M_\odot$ might even be explained by a parent
population entirely older that $\tau\approx 0.3$, scattered by random
errors, whereas a similar interpretation is unlikely for stars with 
$2\,M_\odot < M \le 3\,M_\odot$. In the latter mass range, young
magnetic stars are found more rarely than expected from a distribution
that is homogeneous in age, but they do exist. There are also strong
indications of a lack of lower mass stars in the final stages of the
main sequence evolution. Finally, magnetic CP stars with $M > 3\,M_\odot$ 
are homogeneously distributed in age.

Detection of the Zeeman resolved split lines was reported for 26 
stars included in our sample. They all have masses below $3\,M_\odot$ and
constitute only 13\,\% of the whole sample, which should be compared with 
nearly 70\,\% of such objects in the sample analysed by Hubrig et al. (\cite{HNM00}). 
In Fig.~\ref{fig7} we compare the age distributions of magnetic stars with and 
without magnetically split lines. It is clear that stars with the Zeeman 
resolved lines are more evolved and may not be representative of the parent
population of magnetic CP stars.

\begin{figure*}[!t]
\figps{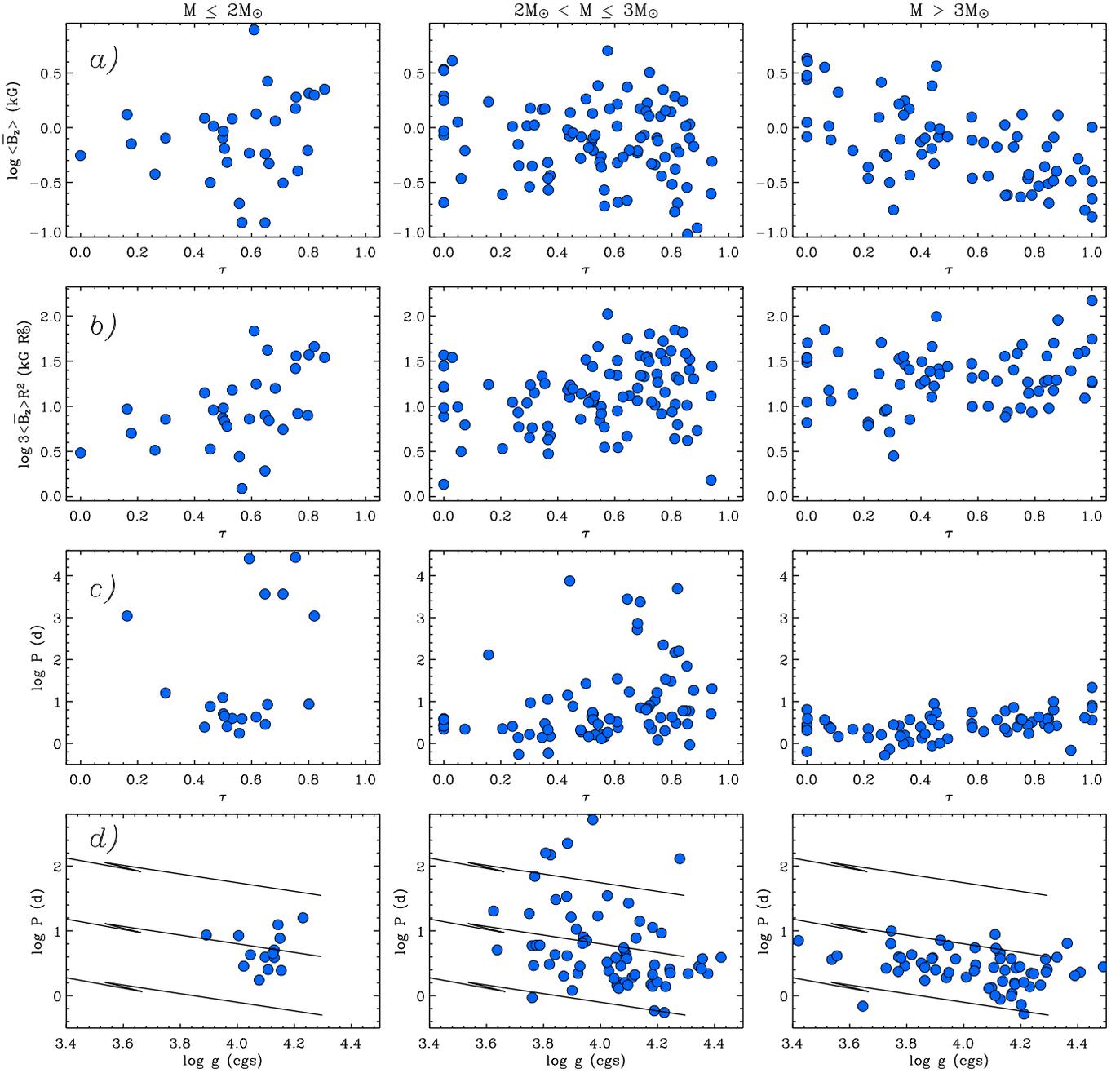}
\caption{The average longitudinal magnetic field (\textbf{a)}), 
magnetic flux (\textbf{b)}), and rotation period (\textbf{c)}) 
as a function of the relative age for magnetic
CP stars with $M\le2\,M_\odot$ (\textit{left column}), 
$2\,M_\odot<M\le3\,M_\odot$ (\textit{middle column}), 
and $M >3\,M_\odot$ (\textit{right column}). The bottom panels (\textbf{d)}) show 
rotation period as a function of surface gravity. The solid lines represent
evolution of the rotation period (for initial periods of 0.5, 4, and 35 days) 
expected for the situation when the angular momentum is conserved.}
\label{fig8}
\end{figure*}

\begin{figure*}[!t]
\figps{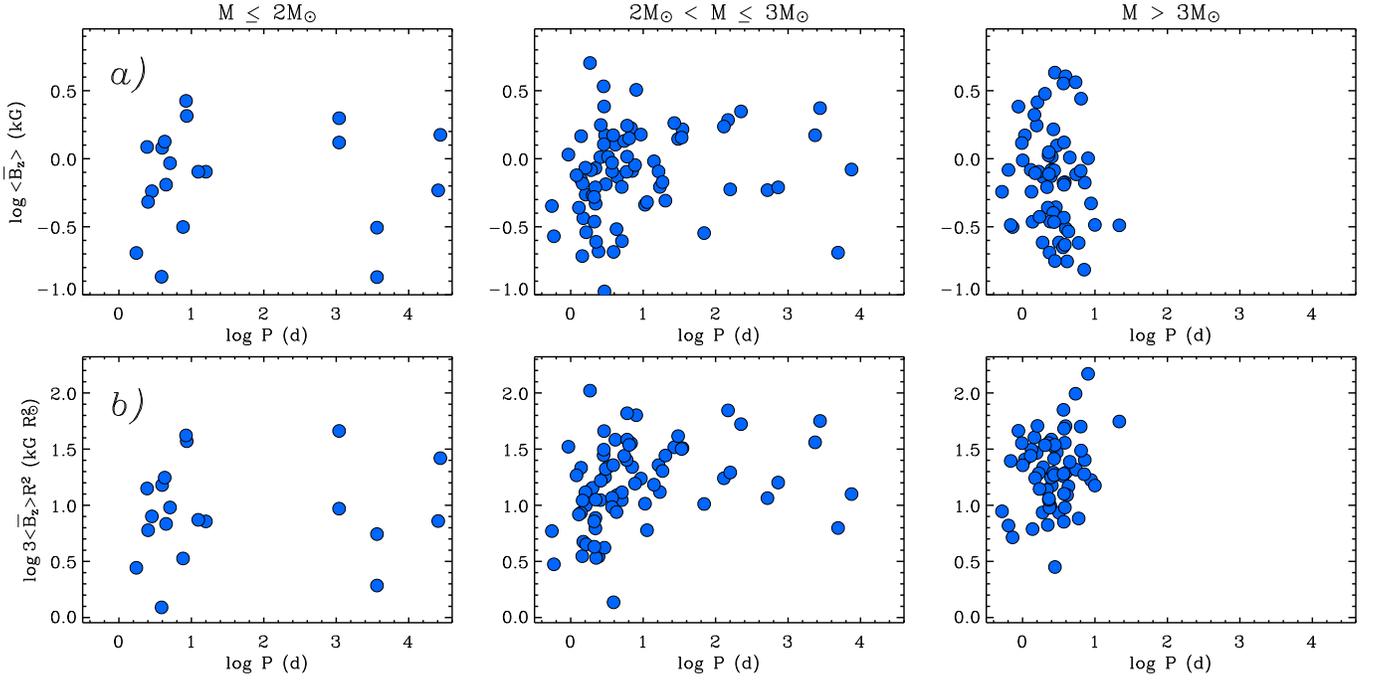}
\caption{The average longitudinal magnetic field (\textbf{a)}) and
magnetic flux (\textbf{b)}) as a function of rotation period.}
\label{fig9}
\end{figure*}

\begin{figure*}[!t]
\figps{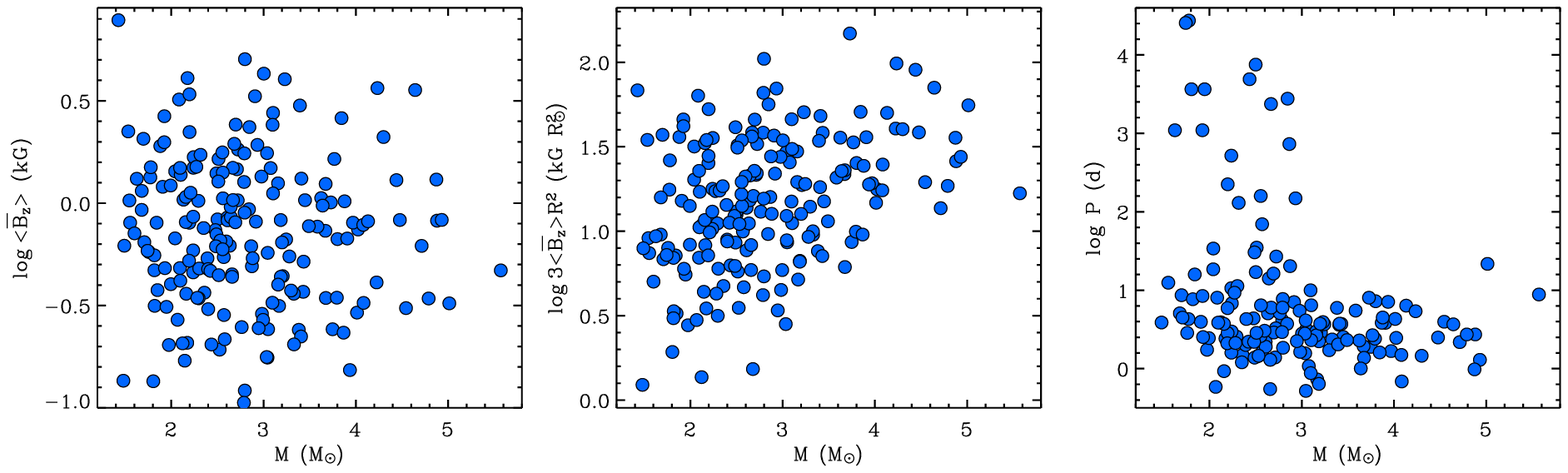}
\caption{The average longitudinal magnetic field (\textit{left panel}), 
magnetic flux (\textit{middle panel}), and rotation period 
(\textit{right panel}) as a function of stellar mass.}
\label{fig10}
\end{figure*}

\subsection{Magnetic field}

Homogeneous determination of the H-R diagram position for the large sample of
magnetic CP stars allows us to investigate evolutionary changes of the surface
magnetic field strength and to probe its possible dependence on the fundamental stellar
parameters. We have considered the average quadratic longitudinal field (Borra et
al. \cite{BLT83}) defined with the equation
\beq
\bBz = \left( \dfrac{1}{N}\sum_{i=1}^N \Bz^2_i\right)^{1/2}
\eeq
as magnetic field strength estimator. When available, the $\bBz$ 
estimates were taken from the catalogue by Bychkov et al. (\cite{BBM03}), 
otherwise we have computed $\bBz$ from the individual longitudinal 
field measurements of the newly detected magnetic CP stars. For a few stars
magnetic field was only observed using Zeeman resolved lines in the intensity spectra.
In this case we have used approximate relation $\bBz\approx
\langle B_{\rm s}\rangle/3$ to bring these field modulus magnetic measurements on 
the same scale with the average longitudinal field estimates. In addition to the
observed magnetic field strength, we have computed the quantity
$3\bBz R^2$, which is proportional to the unsigned magnetic flux and hence
provides a possibility to distinguish intrinsic evolutionary changes of the magnetic
field intensity from the secular variation of the surface field strength caused by increase
in the stellar radii.

The average longitudinal field and magnetic flux as a function of the relative
stellar age $\tau$ are presented in Fig.~\ref{fig8}a and b, where different panels
correspond to the three mass bins defined above. Dependence of the magnetic
quantities on the stellar rotation period and mass are shown in Fig.~\ref{fig9}
and \ref{fig10}, respectively. The non-parametric Spearman's rank correlation
coefficient $r$ (Press et al. \cite{PTV92}) and the associated significance $D$
were used to quantify dependencies of the magnetic quantities on the stellar
parameters.

The plot of $\bBz$ against elapsed fraction of the main sequence life reveals
a significant anticorrelation ($r=-0.29$, $D>99.9$\,\%) for the whole stellar sample
and for the individual groups of stars, except for the low mass ($M\le2M_\odot$)
objects. However, when the strong evolutionary increase of the stellar radii is
accounted for, significant positive correlation ($r=0.24$, $D>99.9$\,\%) between
the magnetic flux and $\tau$ becomes evident. This trend is significant in the
group of $M\le3M_\odot$ stars and marginal for stars with $M>3M_\odot$. On the
basis of this analysis we can draw the conclusion that  the magnetic flux in the
surface layers of low mass ($M\le3M_\odot$) CP stars increases with time. On
average, the magnetic flux grows by almost a factor of 4 between $\tau=0$ and
$\tau=1$. This effect is far less prominent (flux increase of about 40\,\% at the
$D=75$\,\% significance level) in the magnetic stars with $M>3M_\odot$. 

No trend of the average longitudinal field with the stellar mass is present in our
data (Fig.~\ref{fig10}). At the same time, unambiguous correlation ($r=0.35$,
$D>99.9$\,\%) emerges if we consider the magnetic flux as a function of mass. We
have determined the average magnetic fluxes $15.2\pm 0.5$, $19.3\pm 0.2$, and
$27.2\pm 0.3$ (in the $4\pi$kG\,$R^2_\odot$ units) for the $M\le2M_\odot$,
$2M_\odot<M\le3M_\odot$, and $M>3M_\odot$ mass ranges, respectively. Thus, our
investigation leads to the conclusion that massive stars are intrinsically more
magnetic compared to low mass CP objects.

Information on the rotation periods of the magnetic stars in our sample
was obtained from the catalogue of Catalano \& Renson (\cite{CR98}) and 
was further supplemented with the period measurements reported in several
recent studies (Paunzen \& Maitzen \cite{PM98}; Koen \& Eyer \cite{KE02};
Ryabchikova et al. \cite{RWA05}). 
Estimates of rotation periods could be found for 80\,\% of
stars from our sample. A relation between the average longitudinal field and
stellar rotation (see Fig.~\ref{fig9}) reveals a marginal ($D=86$\,\%)
correlation, which is the strongest ($D=98$\,\%) in the group of stars
with $2M_\odot<M\le3M_\odot$. This trend is reinforced if the magnetic flux is
considered instead of the average longitudinal field. In this case, we
have found $r=0.19$, $D=98$\,\% for the whole sample and a definite
correlation ($D>99.9$\,\%) for stars in the intermediate mass range. These
results indicate that the surface field is more intense in slowly
rotating magnetic CP stars.

\subsection{Stellar rotation}

We have investigated evolutionary changes of the stellar rotation periods and
studied a relation between rotation and fundamental stellar parameters. Rotation
period as a function of the relative age, surface gravity, and stellar mass is presented
in Fig.~\ref{fig8}c, Fig.~\ref{fig8}d, and Fig.~\ref{fig10}, respectively. We have
found a clear evidence that older stars have longer rotation periods. This trend
is significant ($r=0.37$, $D>99.9$\,\%) for the whole sample and for the two
groups of stars with $M>2M_\odot$ ($D>99.8$\,\%). A marginal correlation
($D=80$\,\%) is also found for the low mass objects ($M\le2M_\odot$).

A prominent dependence of rotation period on mass is revealed by our
statistical analysis (see Fig.~\ref{fig10}). Rotation
periods of the most massive ($M>3M_\odot$) magnetic CP stars show a fairly
sharp cutoff at $P_{\rm rot}\approx10$~days. In contrast, there are many
instances of much slower rotation among less massive ($M\le3M_\odot$) CP
stars.

Possible changes of the total angular momentum and secular evolution of
the stellar moment of inertia can both contribute to the observed variation of
rotation periods. In order to isolate a signature of the angular momentum
evolution, we have followed North (\cite{N98}) and studied rotation period
as a function of surface gravity. In Fig.~\ref{fig8}d
the observed relation is compared with the changes of rotation period
predicted for rigidly rotating stars that conserve the total angular
momentum during their life at the main sequence. These theoretical curves,
adopted from North (\cite{N98}), are virtually mass independent and are
plotted in Fig.~\ref{fig8}d for initial periods of 0.5, 4, and 35 days.
From this comparison we have found that the observed period-age dependence
of stars with $M>3M_\odot$ is fully accounted for by the changes of the
moment of inertia. Thus, we see no indication of significant changes in
the angular momentum of the most massive magnetic CP stars. At the same
time, the group of stars in the $2M_\odot<M\le3M_\odot$ mass range shows
an abnormally steep slope in their $\log P_{\rm rot}$--$\log g$ diagram,
suggesting that these stars may have experienced some loss of the angular
momentum during the main sequence evolutionary stage. If we subtract
theoretically expected period trend from our data, a marginal residual
anticorrelation ($r=-0.13$, $D=71$\,\%) of rotation period with respect to
$\log g$ is still present. The large scatter of data points precludes us
from quantifying the angular momentum evolution history of the low mass
magnetic CP stars in more detail.

\section{Discussion and conclusions}
\label{discuss}

We have carried out a detailed statistical investigation of the
evolutionary state of the upper main sequence magnetic CP stars. The
sample of 194 objects has included all CP stars whose magnetic status
could be confirmed with a direct detection of the surface magnetic
field and for which precise parallax is available in the Hipparcos
catalogue. The literature data on the magnetic observations of CP
stars were complemented with the analysis of the archival
spectropolarimetric data acquired with the FORS1 instrument at ESO
VLT. This allowed us to detect magnetic field in 53 A and B-type 
CP stars, of which
only five were previously known to be magnetic. Using the
medium-band (Geneva or Str\"omgren) photometry of the program stars,
we have determined \te\ and then have placed stars on the H-R
diagram. We were able to obtain stellar masses, radii, and ages 
from the comparison of the observed luminosity and
temperature with the predictions of the theoretical evolutionary
tracks. Our investigation has included a comprehensive non-linear
error analysis, that has permitted us to quantify uncertainty of the
derived stellar properties for each period of the main sequence life.
Numerical simulations were applied to assess the properties of the
resulting age distributions of the magnetic CP stars of different
masses. We have also employed correlation analysis to study
dependence of the surface magnetic field, magnetic flux, and rotation
period on the stellar mass and to probe possible evolutionary changes of
the stellar rotation and magnetic field.

The key results of our study can be summarised as follows.
\begin{enumerate} 
\item The most massive magnetic CP stars ($M>3M_\odot$) are
distributed homogeneously in the main sequence band. On the
other hand, stars with $M\le3M_\odot$ show the tendency to cluster in
the middle of the main sequence. The relative shortage of young and
very old stars is especially pronounced in the group of low mass 
($M\le2M_\odot$) stars. This uneven age distribution
cannot be attributed to the effect of random errors in determination 
of the stellar parameters.\\
\item We have found 22 young ($\tau\le0.3$) magnetic stars among the 
objects with $M\le3M_\odot$, thereby rejecting the proposal by Hubrig et
al. (\cite{HNM00}) that all observably magnetic low mass CP stars 
have completed significant fraction of their main sequence evolution. 
At the same time, our data for the least massive ($M\le2M_\odot$)
stars is not inconsistent with a population of stars older than $\tau=0.3$.\\
\item The average surface field observed in magnetic CP stars decreases
with time. However, for stars with $M\le3M_\odot$ this decrease is slower
than the field weakening computed under the assumption of the magnetic flux
conservation. Consequently, we suggest that the surface magnetic flux of
the low mass CP stars increases with time.\\
\item Comparison of the average magnetic fluxes of the CP stars from
different mass ranges shows that massive CP stars are substantially 
more magnetic. At the same time, we found a correlation between
magnetic field and rotation period for the intermediate mass stars.\\
\item Rotation period of magnetic stars increases with time, but for
stars with $M>3M_\odot$ the stellar structure changes influencing the 
stellar moment of inertia can fully account for the observed period
increase. On the other hand, our results do not rule out the possibility that
angular momentum losses occur during the main sequence evolution 
of stars with $2\,M_\odot<M\le3\,M_\odot$.
\end{enumerate} 

The conspicuous inhomogeneous age distribution of the low mass magnetic CP
stars which has emerged from our statistical analysis requires careful
verification and interpretation. Our results raise the question whether the
anomalous H-R diagram distribution is the property of stars with significant
surface magnetic field or it is intrinsic to the whole group of the SrCrEu
and less massive Si-type stars. No significant anomalies in the evolutionary
state of these subclasses of CP stars were reported by Gomez et al.
(\cite{GLG98}). However, the adopted upper threshold of the relative
parallax uncertainty and the methods used in the Gomez et al. study differ
substantially from the analysis procedure employed in the present
investigation. Methodologically more similar analysis by P\"{o}hnl et
al.~(\cite{PPM05}) included only 15 stars with $M\le2\,M_\odot$, and most of
these objects were found in the centre of the main sequence band, very
similar to our results. 

An indirect hint that the SrCrEu and Si chemical peculiarity is closely
related to the presence of magnetic field at the stellar surface comes from
the survey of bright Ap stars carried out by Auri\`{e}re et al.
(\cite{ASW04}). Using sensitive spectropolarimetric field diagnostic
methods, these authors were able to detect magnetic field in essentially
every peculiar star they have observed and have established a possible lower
threshold of $\approx$\,250~G for the strength of the dipolar field
component. In the light of these findings and taking results of our study
into account, an investigation of the evolutionary status of \textit{all}
SrCrEu and Si stars with $M\le3\,M_\odot$ would be of great importance and
could benefit from the upcoming revision of the Hipparcos data reduction
(van Leeuwen \& Fantino \cite{LF05}).

For many decades the problem of the evolution of global magnetic field in
A and B stars was approached in the framework of the analysis of the
Ohmic decay of the poloidal fossil field, assumed to exist in
stellar interior (Cowling \cite{C45}; Moss \cite{DM84}; Landstreet
\cite{L87}). Analytical estimates predict appreciable field decay only
after  $\sim$\,10$^{10}$--10$^{11}$~yr, which exceeds the main sequence
lifetime of even the least massive magnetic CP stars. Consequently, any
observation of the possible secular evolution of magnetic field was often
considered as a challenge for the fossil field theory. However, this
classical assessment may be fundamentally flawed due to neglect of the
toroidal field component, which must exist in the interior of magnetic CP
stars in order to ensure dynamical stability of their global fields
(Prendergast \cite{P56}; Tayler \cite{T80}). 
Recent numerical MHD simulations of the fossil
field dynamics in radiative stellar interiors by Braithwaite \& Spruit
(\cite{BS04}, see also Braithwaite \& Nordlund \cite{BN05}) have
established that instability of the poloidal  field component can be
suppressed by the presence of the interior toroidal magnetic 
component of similar strength. The
diffusive evolution of such twisted fields in the simulations by
Braithwaite \& Nordlund (\cite{BN05}) is determined essentially by the
toroidal field. In their model the total magnetic energy decreases
with time, but the surface field strength is expected to increase, until
the toroidal component emerges at the surface and the field decays rapidly.
The time scale of this process is $\sim$\,$2\times10^{9}$~yr for a
$2\,M_\odot$ star (Braithwaite \& Nordlund \cite{BN05}), but the precise
value is rather uncertain due to a very schematic treatment of  the
atmospheric magnetic reconnection and because of the sensitivity to
the assumed initial field structure. These difficulties notwithstanding, the recent
numerical work has emphasized limitations of the traditional analytical
studies of the global field evolution and suggested that an observation
of the long-term systematic change of the surface field is not
necessarily incompatible with the fossil field hypothesis. Our
observation of the uneven distribution of the low mass magnetic CP stars in
the H-R diagram and of the possible increase in their surface magnetic flux
with time may be plausibly interpreted as a signature of the Ohmic
diffusion of the twisted fossil field. The relative shortage of the old
low mass stars may indicate that in these objects we are observing the
final stage of the field emergence and rapid decay. If this scenario is
correct, the field starts to emerge in stars with $M\la2\,M_\odot$ after
$\sim$\,$4\times10^{8}$~yr of the main sequence evolution and completes
its decay after $\sim$\,10$^{9}$~yr.

The difference in the age distribution and magnetic field properties
of the low and high mass CP stars may be related to their pre-main
sequence (PMS) evolutionary history and, especially, to the behaviour
of the interior and envelope convective zones. One can argue (e.g., Tout et
al. \cite{TWF04}) that a large-scale coherent field typical of
magnetic CP stars can be frozen in, or undergo a slow diffusive
evolution as envisaged by Braithwaite \& Nordlund (\cite{BN05}), in
the fully radiative parts of the stellar interiors. In contrast, the
turbulence in the convective zones leads to rapid reconnection of the
field lines and thus contributes to the dissipation of the magnetic
flux. Consequently, primordial field can survive only in stars that do
not pass through a long fully convective phase during their approach
to the ZAMS and in the subsequent main sequence evolution. Theoretical
calculations of the PMS stellar evolution by Palla \& Stahler 
(\cite{PS93}) have shown that, although the low mass stars are fully
convective during a large fraction of their PMS life, the duration of 
this phase decreases rapidly with an increasing stellar mass. A
$1.5\,M_\odot$ star is expected to spend roughly $10^5$ years in the
fully convective phase, which is already a factor of ten shorter
compared to a solar-type star and is probably too short to destroy the
fossil magnetic field completely. For masses of $\ga$\,$2.4\,M_\odot$
the PMS star is never fully convective, which facilitates survival of
the primordial field. Thus, the observed difference in the age
distribution of the low and high mass magnetic CP stars may reflect
the history of the field dissipation in the convectively unstable
regions of stellar interior. The presence of more extended and 
long-lived PMS convective zones in stars with $M \la
2.4-2.0\,M_\odot$ suggests that these objects are more likely to
possess weak or no field in the outer regions when they reach the
ZAMS. Subsequent increase in the field strength results from the
outward expansion of magnetic field into the newly formed radiative
regions. In contrast to this scenario of the field behaviour in the low
mass stars, the fossil field in more massive magnetic CP stars is
probably not significantly altered by the feeble PMS convective zones
and appears close to the stellar surface even in young stars.

In the context of the study of the origin and evolution of the
magnetic field in CP stars it is helpful also to investigate the
evolution of the stellar angular momentum. Magnetic CP stars have
generally longer rotation periods than normal A and B stars. The bulk
of their rotation rates forms a separate Maxwellian distribution with
an average value 3--4 times lower than that found in normal A and
B-type stars (St\c{e}pie\'{n} \cite{S00}), but there are also groups
of CP stars with rotation period of years (Mathys et al. \cite{MHL97})
or decades (e.g., $\gamma$ Equ, see Leroy et al. \cite{LBL94}).
Several works suggest that neither field CP stars nor cluster CP stars
undergo any significant magnetic braking during their life on the main
sequence (North~\cite{N98}; this study). Therefore, angular momentum
must be lost before the star reaches the ZAMS.  St\c{e}pie\'{n}
(\cite{S00}) explains the slow rotation as the result of an
interaction of the stellar magnetic field with the circumstellar
environment during the PMS phase. And if a magnetised wind still
persists after the dissipation of the circumstellar disk, a PMS CP
star may further slow down, reaching the ZAMS with an extremely long
rotation period. In this respect, the very existence of slow rotating
stars with strong magnetic fields suggests that magnetic field was
already present when the stars were in the pre-main sequence phase.
For stars with $M>2\,M_\odot$ this hypothesis is supported by
the discovery of a magnetic field in NGC\,2244~334, a $4\,M_\odot$
star that has spent only 2\,\% of its life in the main sequence
(Bagnulo et al. \cite{BHL04}), and in HD~66318, a star with a mass of
$2.1\,M_\odot$ and $\tau = 0.16$, that belongs to the open cluster
NGC~2516. More direct confirmation that magnetic fields are present
during the pre-main sequence phase comes from the discovery of magnetic
field in Herbig Ae/Be stars (considered the progenitors of 
main sequence early-type stars) by Wade et al. (\cite{WDB05}).

The most puzzling evolutionary properties are observed for the group
of stars with $M \le 2\,M_\odot$, which contains very few young stars
and no stars approaching the terminal age main sequence.
Observational evidence for lack of young lower mass ($M \la
2\,M_\odot$) Ap stars comes also from open cluster studies. Abt
(\cite{A79}) suggested that low-mass Ap stars are found only in
clusters that are at least $10^8$\,yr old, and a study of four nearby
young clusters by P\"{o}hnl et al. (\cite{PMP03}) is also 
consistent with this finding. Assuming that magnetic breaking during
pre-main sequence phase is the reason why these stars rotate slower
than normal late A-type stars, then we face a scenario in which magnetic
field appears and disappears from the stellar surface several times
during the star life. Magnetic field was present at the surface of the
star during the pre-main sequence phase, then disappeared when the
star reached the ZAMS, to appear again at a more evolved state, and
disappear toward the end of the star's life in the main sequence.
At the same time, one can argue that magnetic field responsible for the PMS angular
momentum loss is not directly related to the fossil field seen at the 
surfaces of main sequence low mass magnetic CP stars. The former field
may be generated by the dynamo processes or represent the outer part of the
fossil field tangled by the envelope convection. This complex field decays by
the time the star reaches the ZAMS and, after some evolution on the main
sequence, the interior fossil field which has retained its global organisation
appears at the surface.

Our study is the first investigation which included enough very low
mass magnetic CP stars to identify interesting characteristics of this
stellar group, yet it is clear that the number of stars in the
corresponding mass range is still rather small. This circumstance does
not permit us to draw definite conclusions about the nature of the
anomalies in the age distribution observed for these stars. Thus, we
call for a more detailed analysis of the rotation, magnetic, and
evolutionary characteristics of CP stars with
$M\le2\,M_\odot$. Such an investigation would represent the most
interesting follow up of the present study.

\begin{acknowledgements}
This research has made extensive use of the {\sc simbad} database, operated
at CDS, Strasbourg, France, and NASA's Astrophysics Data System
Bibliographic Services. OK acknowledges funding from the Scientific
Visitors Programme of ESO Chile. We acknowledge the use of ESO Science
Archive Facility.
\end{acknowledgements}

\Online

{\small
\begin{longtable}{llrlrr@{$\pm$}rc}
\caption{\label{tbl1} \Bz\ measurements for the sample of CP stars
observed with FORS1 (data retrieved from the ESO archive). 
Columns 1 and 2 list the HD and HIP identification. Columns 3
and 4 give the $V$-magnitude and spectral type, respectively, most of 
which are extracted from the \textit{General catalogue of Ap and Am stars} by Renson et
al.\ (\cite{RGC91}). Column~5 gives the Julian Date of the middle of the
exposures. Column~6 reports \Bz\ with its error bar in
Gauss. The last column indicates stars for which magnetic field was 
detected for the first time (new detection, ND) or confirmed 
(confirmed detection, CD).}\\
\hline \hline
HD        &  
HIP       &
$V$       & 
Sp.\ Type &
\multicolumn{1}{c}{JD}  &
\multicolumn{2}{c}{\Bz\ (G)} &
\multicolumn{1}{c}{Comment} \\
\hline
\endfirsthead
\caption{Continued.}\\
\hline \hline
HD        &  
HIP       &
$V$       & 
Sp.\ Type &
\multicolumn{1}{c}{JD}  &
\multicolumn{2}{c}{\Bz\ (G)} &
\multicolumn{1}{c}{Comment} \\
\hline
\endhead
\hline
\endfoot
1048   &  1193&6.2 &A1 Si          & 2452910.603&   403  &  105 &ND \\  
       &      &    &               & 2453199.906&  $-$89 &   46 &   \\  
       &      &    &               & 2453215.882&     36 &   45 &   \\  
3326   &  2852&6.1 &A6 Sr          & 2452908.690&    99  &   62 &   \\  
8783   &  6534&7.8 &A2 Sr\,Cr\,Eu  & 2452852.858& $-$29  &  106 &   \\  
10840  &  8132&6.8 &B9 Si          & 2453184.831& $-$148 &   81 &   \\  
19712  & 14736&7.3 &A0 Cr\,Eu      & 2452905.884&$-$930  &  109 &ND \\  
       &      &    &               & 2452999.525&   764  &   66 &   \\  
19918  & 14026&9.4 &A5 Sr\,Cr\,Eu  & 2452908.711&$-$777  &  109 &ND \\  
22374  & 16859&6.7 &A1 Cr\,Sr\,Si  & 2452999.538& $-$10  &   64 &   \\  
       &      &    &               & 2453216.880&     63 &   36 &   \\  
22488  & 16527&7.7 &A3 Sr\,Cr\,Eu  & 2453087.514&    114 &   51 &   \\  
23207  & 17345&7.5 &A2 Sr\,Eu      & 2453215.861&    259 &   72 &ND \\  
       &      &    &               & 2453218.835&    394 &   53 &   \\  
23408  & 17573&3.9 &B7 He-weak\,Mn & 2452963.656&    24  &   65 &   \\  
24188  & 17543&6.3 &A0 Si          & 2453087.532&    426 &   48 &ND \\  
30612  & 21949&5.5 &B9 Si          & 2453087.546&     44 &   44 &   \\  
34797  & 24827&6.5 &B8 He-weak\,Si & 2452999.566&  1059  &   86 &ND \\  
42659  & 29365&6.7 &A3 Sr\,Cr\,Eu  & 2452999.619&   418  &   78 &ND \\  
55522  & 34798&5.9 &B2 Si\,He      & 2452999.699&   168  &   81 &   \\  
       &      &    &               & 2453000.553&   821  &   68 &ND \\  
56350  & 34929&6.7 &A0 Sr\,Cr\,Eu  & 2452999.739&   824  &   86 &ND \\  
56455  & 35029&5.7 &A0 Si          & 2452999.751&    85  &   98 &   \\  
58448  & 35676&7.1 &B8 Si          & 2452999.765&    60  &   99 &   \\  
60435  & 36537&8.9 &A3 Sr\,Eu      & 2453000.572&$-$315  &   87 &ND \\  
63401  & 37982&6.3 &B9 Si          & 2453002.553&    48  &  113 &   \\  
       &      &    &               & 2453004.728&$-$486  &  106 &ND \\  
74168  & 42519&7.5 &B9 Si          & 2453002.611& $-$75  &   76 &   \\  
74196  & 42535&5.6 &B7 He-weak     & 2452906.888&   121  &  122 &   \\  
75989  & 43528&6.5 &B9 Si          & 2452992.841&$-$279  &  161 &   \\  
       &      &    &               & 2453004.783& $-$46  &  102 &   \\  
80316  & 45658&7.8 &A3 Sr\,Eu      & 2452992.857&$-$269  &  134 &   \\  
83625  & 47272&6.9 &A0 Si\,Sr      & 2453008.822&$-$1484 &   83 &ND \\  
84041  &      &9.4 &A5 Sr\,Eu      & 2453002.670&   497  &   91 &ND \\  
86181  & 48619&9.4 &F0 Sr          & 2453002.692&   360  &   77 &ND \\  
86199  & 48643&6.7 &B9 Si          & 2453003.845&$-$768  &   89 &ND \\  
88158  & 49642&6.5 &B8 Si          & 2453008.838&   233  &   77 &ND \\  
88385  & 49791&8.1 &A0 Si\,Cr\,Eu  & 2453010.681&$-$958  &   70 &ND \\  
89103  & 50248&7.8 &B9 Si          & 2453010.702&$-$1949 &   84 &ND \\  
89385  & 50398&8.4 &B9 Si\,Cr\,Eu  & 2453010.718& $-$81  &   87 &   \\  
91239  & 51512&7.4 &B9 Si\,Cr\,Eu  & 2453118.559& $-$116 &   77 &   \\  
92106  & 51632&7.8 &A0 Sr\,Cr\,Eu  & 2453010.739&    72  &   98 &   \\  
       &      &    &               & 2453118.580&  $-$89 &   90 &   \\  
92385  & 52059&6.7 &B9 Si          & 2453008.869&$-$588  &   90 &ND \\  
       &      &    &               & 2453020.832&   240  &  101 &   \\  
92499  & 52218&8.9 &A2 Sr\,Cr\,Eu  & 2453010.755&$-$1163 &  324 &ND \\  
       &      &    &               & 2453011.712&$-$1230 &  124 &   \\  
       &      &    &               & 2453118.595& $-$989 &  141 &   \\  
93030  & 52419&2.7 &B0 Si\,N\,P    & 2453012.731& $-$46  &  132 &   \\  
96451  & 54166&6.9 &A0 Sr          & 2453074.840&      1 &   51 &   \\  
98340  & 55181&7.1 &B9 Si          & 2453074.862&   1033 &   65 &ND \\  
99563  & 55890&8.5 &F0 Sr          & 2453012.747&$-$392  &  124 &CD \\  
       &      &    &               & 2453015.725&$-$669  &  159 &   \\  
105379 & 59167&8.0 &A0 Sr\,Cr      & 2453011.750&    25  &   83 &   \\  
105382 & 59173&4.5 &B6 He          & 2453011.695&$-$1000 &  101 &ND \\  
       &      &    &               & 2453015.746&$-$610  &  136 &   \\  
105770 & 59404&7.4 &B9 Si          & 2453011.733&   449  &  110 &ND \\  
       &      &    &               & 2453120.645&    262 &   71 &   \\  
105999 & 59487&7.4 &F1 Sr\,Cr      & 2453011.770& $-$30  &  128 &   \\  
107696 & 60379&5.4 &B8 Cr          & 2452824.530& $-$46  &  108 &   \\  
       &      &    &               & 2453074.875&  $-$75 &  142 &   \\  
108945 & 61071&5.5 &A3 Sr          & 2453015.835&    65  &  116 &   \\  
114365 & 64320&6.1 &A0 Si          & 2452824.543&     8  &  100 &   \\  
115226 & 64883&8.5 &A3 Sr          & 2453086.799&    677 &   57 &ND \\  
115440 & 65053&8.2 &B9 Si          & 2453077.714&   3217 &   61 &ND \\  
116890 & 65755&6.2 &B9 Si          & 2452824.555&$-$292  &   77 &ND \\  
117025 & 65783&6.1 &A2 Sr\,Cr\,Eu  & 2452824.567&   483  &   84 &ND \\  
       &      &    &               & 2453120.664&    463 &   68 &   \\  
118913 & 66888&7.7 &A0 Sr\,Cr\,Eu  & 2452824.581& $-$345 &   88 &ND \\  
       &      &    &               & 2453120.681& $-$555 &   38 &   \\  
119308 & 66942&7.8 &A0 Sr\,Cr\,Eu  & 2453120.704& $-$326 &   61 &ND \\  
122970 & 68790&8.3 &F0 Sr\,Cr\,Eu  & 2453015.850&   526  &  137 &CD \\  
125630 & 70346&6.8 &A2 Si\,Cr\,Sr  & 2452824.607&   660  &   67 &ND \\  
       &      &    &               & 2453120.721&     30 &   55 &   \\  
127453 & 71314&7.4 &B8 Si          & 2452824.621&$-$361  &   85 &ND \\  
127575 & 71359&7.7 &B9 Si          & 2453079.888&    911 &   64 &ND \\  
128775 & 71727&6.6 &B9 Si          & 2453120.736& $-$278 &   52 &ND \\  
128974 & 71783&5.7 &A0 Si          & 2452824.644& $-$43  &   55 &   \\  
129899 & 72670&6.4 &A0 Si          & 2453120.795&    495 &   42 &ND \\  
130158 & 72323&5.6 &B9 Si          & 2452824.676&     2  &   53 &   \\  
       &      &    &               & 2453116.812&      9 &   45 &   \\  
130557 & 72449&6.1 &B9 Si\,Cr      & 2452853.558& $-$10  &   70 &   \\  
       &      &    &               & 2453144.767&     19 &   43 &   \\  
131120 & 72800&5.0 &B7 He-weak     & 2452824.660&$-$152  &  114 &   \\  
       &      &    &               & 2453020.857&    57  &   77 &   \\  
       &      &    &               & 2453030.864&   152  &  125 &   \\  
132322 & 73520&7.4 &A7 Sr\,Cr\,Eu  & 2453111.811&    340 &   40 &ND \\  
133792 & 74181&6.3 &A0 Sr\,Cr      & 2452853.570&    119 &   77 &CD \\  
       &      &    &               & 2453120.812&    124 &   40 &   \\  
134305 & 74109&7.2 &A6 Sr\,Cr\,Eu  & 2453144.801&    170 &   49 &ND \\  
136933 & 75439&5.4 &A0 Si          & 2452823.720&     23 &   89 &   \\  
138758 & 76767&7.9 &B9 Si          & 2453086.828&    430 &   41 &ND \\  
138764 & 76243&5.2 &B6 Si          & 2452904.515&   202  &   95 &   \\  
138769 & 76371&4.5 &B3 He          & 2452904.503&    91  &   95 &   \\  
       &      &    &               & 2452904.527&   123  &   91 &   \\  
       &      &    &               & 2452908.522&$-$166  &  120 &   \\  
145102 & 79235&6.6 &B9 Si          & 2452763.815&    40  &   79 &   \\  
147869 & 80351&5.8 &A1 Sr          & 2452763.827&    53  &   74 &   \\  
       &      &    &               & 2453144.818&      7 &   40 &   \\  
148112 & 80463&4.6 &A0 Cr\,Eu      & 2452763.838& $-$59  &   64 &   \\  
148898 & 80975&4.4 &A6 Sr\,Cr\,Eu  & 2452763.849&   241  &   84 &   \\  
149764 & 81477&6.9 &A0 Si          & 2452763.874&$-$1169 &   86 &ND \\  
       &      &    &               & 2453120.831&     49 &   48 &   \\  
149822 & 81337&6.4 &B9 Si\,Cr      & 2452763.861&$-$657  &   66 &ND \\  
150549 & 82129&5.1 &A0 Si          & 2452763.886&$-$167  &   63 &   \\  
       &      &    &               & 2453116.886&  $-$52 &   60 &   \\  
       &      &    &               & 2453120.850&  $-$49 &   34 &   \\  
151525 & 82216&5.2 &B9 Eu\,Cr      & 2452733.895&    76  &   73 &   \\  
       &      &    &               & 2452763.897&   237  &   75 &CD \\  
154708 & 84017&8.8 &A2 Sr\,Cr\,Eu  & 2453120.876&   6859 &   58 &CD \\  
157751 & 85372&7.6 &B9 Si\,Cr      & 2452793.771&  4070  &   65 &ND \\  
       &      &    &               & 2453116.904&   3982 &   48 &   \\  
160468 & 86930&7.3 &F2 Sr\,Cr      & 2453116.862&  $-$96 &   83 &   \\  
       &      &    &               & 2453134.819&  $-$55 &   54 &   \\  
161277 & 86983&7.1 &B9 Si          & 2453134.840&     94 &   44 &   \\  
166469 & 89178&6.5 &A0 Si\,Cr\,Sr  & 2452793.791& $-$42  &   49 &   \\  
       &      &    &               & 2453136.772&  $-$26 &   45 &   \\  
168856 & 90030&7.0 &B9 Si          & 2453144.840& $-$530 &   59 &ND \\  
171184 & 91001&8.0 &A0 Si          & 2452880.529&    250 &   52 &ND \\  
       &      &    &               & 2453144.868&  $-$14 &   48 &   \\  
171279 & 91031&7.3 &A0 Sr\,Cr\,Eu  & 2453144.893&  $-$40 &   40 &   \\  
172032 & 91414&7.7 &A9 Sr\,Cr      & 2453151.605&  $-$31 &   55 &   \\  
172690 & 93481&7.5 &A0 Si\,Sr\,Cr  & 2452793.814&$-$287  &   86 &ND \\  
       &      &    &               & 2453134.868&    235 &   52 &   \\  
175744 & 92934&6.6 &B9 Si          & 2452880.555&   104  &   76 &   \\  
       &      &    &               & 2452901.519&   162  &   91 &   \\  
176196 & 93863&7.5 &B9 Eu\,Cr      & 2452793.829&   240  &   83 &ND \\  
       &      &    &               & 2453134.889&    190 &   51 &   \\  
183806 & 96178&5.6 &A0 Cr\,Eu\,Sr  & 2452793.845& $-$23  &   64 &ND \\  
       &      &    &               & 2453120.924&    148 &   37 &   \\  
186117 & 97533&7.3 &A0 Sr\,Cr\,Eu  & 2453134.913&  $-$19 &   49 &   \\  
       &      &    &               & 2453140.829&     27 &   46 &   \\  
192674 &100090&7.5 &B9 Cr\,Eu\,Sr  & 2453137.861&      7 &   44 &   \\  
199180 &103246&7.7 &A0 Si\,Cr      & 2452822.844&$-$228  &   85 &   \\  
199728 &103616&6.2 &B9 Si          & 2452822.857&$-$245  &   73 &ND \\  
201018 &104337&8.6 &A2 Cr\,Eu      & 2453151.871&    546 &   43 &ND \\  
202627 &105140&4.7 &A1 Si          & 2452793.874& $-$56  &   68 &   \\  
206653 &107525&7.2 &B9 Si          & 2452793.894&    32  &   68 &   \\  
212385 &110624&6.8 &A3 Sr\,Cr\,Eu  & 2452822.913&   163  &   72 &ND \\  
       &      &    &               & 2453184.797&    626 &   52 &   \\  
221760 &116389&4.7 &A2 Sr\,Cr\,Eu  & 2452793.915& $-$48  &   97 &   \\  
       &      &    &               & 2453184.814&     62 &   65 &   \\  
\hline                               
\end{longtable}   
}                   

\newpage

{\small
\begin{longtable}{rrrcccccc}
\caption{\label{tbl2} Fundamental parameters of magnetic CP stars. The columns give numbers in the HD and
Hipparcos catalogues, distance determined from the Hipparcos parallax, absolute magnitude, \te, luminosity 
and mass in solar units, absolute and fractional
stellar age. For the last two columns numbers in brackets give 1$\sigma$ ranges compatible with the errors
of \te\ and $L/L_{\sun}$.}\\
\hline
\hline
 HD~~ & HIP~ & d (pc)~ & $M_V$ & $\log\,T_{\rm eff}$ (K) & $\log\,L/L_{\sun}$ & $M/M_{\sun}$ & $\log\,t$ (yr) & $\tau$ \\ 
\hline
\endfirsthead
\caption{Continued.}\\
\hline
\hline
 HD~~ & HIP~ & d (pc)~ & $M_V$ & $\log\,T_{\rm eff}$ (K) & $\log\,L/L_{\sun}$ & $M/M_{\sun}$ & $\log\,t$ (yr) & $\tau$ \\ 
\hline
\endhead
\hline
\endfoot
   1048 &   1193 & 108$\pm$8~~ & \phantom{$-$}1.05$\pm$0.17 &  3.949$\pm$0.015 & 1.50$\pm$0.07 & 2.17$\pm$0.06 &8.72 (8.63$-$8.79) & 0.61 (0.49$-$0.71) \\
   2453 &   2243 & 151$\pm$18  & \phantom{$-$}0.88$\pm$0.27 &  3.949$\pm$0.015 & 1.57$\pm$0.11 & 2.24$\pm$0.10 &8.72 (8.65$-$8.78) & 0.68 (0.54$-$0.78) \\
   3980 &   3277 &  65$\pm$2~~ & \phantom{$-$}1.62$\pm$0.09 &  3.917$\pm$0.011 & 1.24$\pm$0.04 & 1.91$\pm$0.03 &8.83 (8.75$-$8.89) & 0.53 (0.45$-$0.61) \\
   4778 &   3919 &  90$\pm$6~~ & \phantom{$-$}1.23$\pm$0.15 &  3.999$\pm$0.013 & 1.51$\pm$0.07 & 2.29$\pm$0.06 &8.26 (7.53$-$8.47) & 0.24 (0.03$-$0.40) \\
   5737 &   4577 & 206$\pm$35  &           $-$2.31$\pm$0.38 &  4.121$\pm$0.013 & 3.19$\pm$0.15 & 5.01$\pm$0.21 &7.97 (7.92$-$8.01) & 1.00 (1.00$-$1.00) \\
   8441 &   6560 & 203$\pm$33  & \phantom{$-$}0.17$\pm$0.35 &  3.956$\pm$0.014 & 1.86$\pm$0.14 & 2.57$\pm$0.18 &8.66 (8.61$-$8.71) & 0.85 (0.76$-$0.93) \\
   9996 &   7651 & 139$\pm$16  & \phantom{$-$}0.80$\pm$0.26 &  4.012$\pm$0.013 & 1.70$\pm$0.11 & 2.50$\pm$0.10 &8.41 (8.16$-$8.51) & 0.44 (0.23$-$0.60) \\
  10221 &   7965 & 136$\pm$11  &           $-$0.21$\pm$0.18 &  4.030$\pm$0.016 & 2.15$\pm$0.08 & 3.05$\pm$0.11 &8.42 (8.37$-$8.46) & 0.79 (0.69$-$0.86) \\
  10783 &   8210 & 186$\pm$25  & \phantom{$-$}0.08$\pm$0.30 &  4.006$\pm$0.013 & 1.98$\pm$0.12 & 2.79$\pm$0.15 &8.51 (8.47$-$8.55) & 0.76 (0.65$-$0.85) \\
  11187 &   8643 & 234$\pm$44  & \phantom{$-$}0.14$\pm$0.41 &  4.029$\pm$0.016 & 2.00$\pm$0.17 & 2.87$\pm$0.20 &8.43 (8.33$-$8.49) & 0.68 (0.48$-$0.82) \\
  11503 &   8832 &  62$\pm$3~~ & \phantom{$-$}0.60$\pm$0.12 &  4.010$\pm$0.013 & 1.78$\pm$0.05 & 2.57$\pm$0.06 &8.47 (8.38$-$8.54) & 0.55 (0.44$-$0.65) \\
  12288 &   9604 & 230$\pm$38  & \phantom{$-$}0.59$\pm$0.36 &  3.996$\pm$0.013 & 1.75$\pm$0.15 & 2.51$\pm$0.15 &8.54 (8.42$-$8.60) & 0.61 (0.42$-$0.75) \\
  12447 &   9487 &  42$\pm$1~~ & \phantom{$-$}1.03$\pm$0.10 &  3.999$\pm$0.013 & 1.58$\pm$0.05 & 2.36$\pm$0.05 &8.41 (8.21$-$8.53) & 0.37 (0.23$-$0.49) \\
  12767 &   9677 & 110$\pm$9~~ &           $-$0.56$\pm$0.19 &  4.111$\pm$0.013 & 2.47$\pm$0.08 & 3.75$\pm$0.13 &8.13 (8.07$-$8.18) & 0.70 (0.59$-$0.79) \\
  14437 &  10951 & 197$\pm$36  & \phantom{$-$}0.51$\pm$0.40 &  4.034$\pm$0.016 & 1.87$\pm$0.16 & 2.72$\pm$0.17 &8.36 (8.03$-$8.46) & 0.50 (0.21$-$0.69) \\
  15089 &  11569 &  43$\pm$1~~ & \phantom{$-$}1.46$\pm$0.08 &  3.925$\pm$0.010 & 1.31$\pm$0.03 & 1.97$\pm$0.03 &8.80 (8.74$-$8.86) & 0.56 (0.48$-$0.62) \\
  15144 &  11348 &  65$\pm$4~~ & \phantom{$-$}1.91$\pm$0.14 &  3.926$\pm$0.010 & 1.13$\pm$0.06 & 1.84$\pm$0.04 &8.63 (8.30$-$8.77) & 0.30 (0.13$-$0.42) \\
  17775 &  13507 & 156$\pm$23  & \phantom{$-$}1.92$\pm$0.33 &  3.930$\pm$0.015 & 1.13$\pm$0.13 & 1.85$\pm$0.09 &8.57 (7.05$-$8.81) & 0.26 (0.00$-$0.50) \\
  18296 &  13775 & 118$\pm$12  &           $-$0.44$\pm$0.23 &  4.036$\pm$0.016 & 2.25$\pm$0.10 & 3.21$\pm$0.15 &8.38 (8.34$-$8.43) & 0.83 (0.74$-$0.90) \\
  18610 &  13534 & 202$\pm$27  & \phantom{$-$}1.55$\pm$0.30 &  3.878$\pm$0.012 & 1.27$\pm$0.12 & 1.88$\pm$0.11 &9.00 (8.96$-$9.04) & 0.76 (0.65$-$0.84) \\
  19712 &  14376 & 166$\pm$25  & \phantom{$-$}1.16$\pm$0.33 &  4.056$\pm$0.015 & 1.66$\pm$0.14 & 2.61$\pm$0.11 &6.74 (6.65$-$8.04) & 0.00 (0.00$-$0.22) \\
  19805 &  14980 & 168$\pm$24  & \phantom{$-$}1.41$\pm$0.32 &  3.973$\pm$0.014 & 1.39$\pm$0.13 & 2.13$\pm$0.10 &8.43 (7.05$-$8.63) & 0.29 (0.00$-$0.51) \\
  19832 &  14893 & 113$\pm$11  & \phantom{$-$}0.35$\pm$0.22 &  4.095$\pm$0.014 & 2.07$\pm$0.09 & 3.17$\pm$0.12 &7.95 (7.33$-$8.16) & 0.29 (0.06$-$0.49) \\
  21699 &  16470 & 179$\pm$22  &           $-$1.05$\pm$0.28 &  4.159$\pm$0.012 & 2.78$\pm$0.12 & 4.48$\pm$0.22 &7.96 (7.91$-$8.01) & 0.74 (0.62$-$0.83) \\
  22316 &  16974 & 170$\pm$20  & \phantom{$-$}0.04$\pm$0.26 &  4.073$\pm$0.015 & 2.14$\pm$0.11 & 3.16$\pm$0.13 &8.24 (8.09$-$8.32) & 0.58 (0.39$-$0.71) \\
  22374 &  16859 & 134$\pm$17  & \phantom{$-$}0.82$\pm$0.28 &  3.938$\pm$0.015 & 1.58$\pm$0.11 & 2.24$\pm$0.12 &8.76 (8.71$-$8.81) & 0.74 (0.62$-$0.83) \\
  22470 &  16803 & 145$\pm$17  &           $-$0.38$\pm$0.26 &  4.115$\pm$0.013 & 2.41$\pm$0.11 & 3.67$\pm$0.15 &8.10 (7.99$-$8.16) & 0.62 (0.46$-$0.74) \\
  22920 &  17167 & 226$\pm$38  &           $-$1.32$\pm$0.37 &  4.142$\pm$0.013 & 2.85$\pm$0.15 & 4.54$\pm$0.32 &8.01 (7.96$-$8.05) & 0.85 (0.74$-$0.93) \\
  23207 &  17345 & 177$\pm$31  & \phantom{$-$}1.28$\pm$0.38 &  3.896$\pm$0.011 & 1.38$\pm$0.15 & 2.00$\pm$0.14 &8.93 (8.87$-$8.97) & 0.76 (0.64$-$0.85) \\
  23408 &  17573 & 110$\pm$12  &           $-$1.54$\pm$0.25 &  4.079$\pm$0.014 & 2.79$\pm$0.10 & 4.22$\pm$0.18 &8.15 (8.12$-$8.18) & 0.97 (0.92$-$1.00) \\
  24155 &  18033 & 135$\pm$16  & \phantom{$-$}0.32$\pm$0.27 &  4.132$\pm$0.013 & 2.17$\pm$0.11 & 3.45$\pm$0.14 &7.30 (6.27$-$7.88) & 0.08 (0.00$-$0.32) \\
  24188 &  17543 & 142$\pm$10  & \phantom{$-$}0.40$\pm$0.17 &  4.101$\pm$0.014 & 2.06$\pm$0.07 & 3.19$\pm$0.10 &7.81 (6.67$-$8.07) & 0.21 (0.01$-$0.40) \\
  24712 &  18339 &  48$\pm$2~~ & \phantom{$-$}2.54$\pm$0.09 &  3.857$\pm$0.012 & 0.87$\pm$0.04 & 1.55$\pm$0.03 &9.07 (8.94$-$9.17) & 0.50 (0.37$-$0.61) \\
  25267 &  18673 & 101$\pm$7~~ &           $-$0.30$\pm$0.15 &  4.080$\pm$0.014 & 2.30$\pm$0.07 & 3.38$\pm$0.10 &8.24 (8.18$-$8.29) & 0.70 (0.59$-$0.78) \\
  25354 &  18912 & 144$\pm$20  & \phantom{$-$}1.79$\pm$0.31 &  3.993$\pm$0.013 & 1.27$\pm$0.13 & 2.12$\pm$0.08 &7.05 (7.02$-$8.05) & 0.00 (0.00$-$0.11) \\
  25823 &  19171 & 151$\pm$19  &           $-$0.63$\pm$0.28 &  4.112$\pm$0.013 & 2.50$\pm$0.12 & 3.80$\pm$0.19 &8.13 (8.07$-$8.17) & 0.72 (0.59$-$0.82) \\
  27309 &  20186 &  96$\pm$7~~ & \phantom{$-$}0.40$\pm$0.16 &  4.079$\pm$0.014 & 2.01$\pm$0.07 & 3.04$\pm$0.09 &8.07 (7.73$-$8.23) & 0.34 (0.15$-$0.51) \\
  28843 &  21192 & 131$\pm$14  & \phantom{$-$}0.06$\pm$0.24 &  4.143$\pm$0.013 & 2.30$\pm$0.10 & 3.67$\pm$0.14 &7.65 (6.26$-$7.93) & 0.21 (0.00$-$0.43) \\
  30466 &  22402 & 163$\pm$25  & \phantom{$-$}0.68$\pm$0.34 &  4.044$\pm$0.016 & 1.82$\pm$0.14 & 2.71$\pm$0.14 &8.21 (7.40$-$8.40) & 0.34 (0.04$-$0.57) \\
  32633 &  23733 & 156$\pm$22  & \phantom{$-$}0.72$\pm$0.32 &  4.108$\pm$0.014 & 1.95$\pm$0.13 & 3.10$\pm$0.13 &6.39 (6.33$-$7.82) & 0.00 (0.00$-$0.21) \\
  34452 &  24799 & 137$\pm$13  &           $-$0.33$\pm$0.21 &  4.160$\pm$0.012 & 2.49$\pm$0.09 & 4.02$\pm$0.14 &7.82 (7.51$-$7.94) & 0.40 (0.19$-$0.55) \\
  34797 &  24827 & 238$\pm$47  &           $-$0.47$\pm$0.43 &  4.102$\pm$0.014 & 2.41$\pm$0.17 & 3.62$\pm$0.26 &8.16 (8.05$-$8.21) & 0.69 (0.49$-$0.82) \\
  38823 &  27423 & 113$\pm$12  & \phantom{$-$}2.00$\pm$0.24 &  3.839$\pm$0.013 & 1.09$\pm$0.10 & 1.70$\pm$0.08 &9.16 (9.11$-$9.21) & 0.80 (0.71$-$0.88) \\
  39317 &  27743 & 157$\pm$22  &           $-$0.50$\pm$0.31 &  4.018$\pm$0.013 & 2.24$\pm$0.13 & 3.16$\pm$0.19 &8.42 (8.37$-$8.47) & 0.88 (0.80$-$0.93) \\
  40312 &  28380 &  53$\pm$2~~ &           $-$1.01$\pm$0.10 &  4.007$\pm$0.013 & 2.42$\pm$0.05 & 3.41$\pm$0.07 &8.36 (8.34$-$8.39) & 1.00 (0.94$-$1.00) \\
  42616 &  29565 & 173$\pm$28  & \phantom{$-$}0.56$\pm$0.35 &  3.989$\pm$0.013 & 1.76$\pm$0.14 & 2.50$\pm$0.15 &8.57 (8.49$-$8.62) & 0.65 (0.48$-$0.77) \\
  42659 &  29365 & 135$\pm$14  & \phantom{$-$}1.03$\pm$0.23 &  3.900$\pm$0.011 & 1.48$\pm$0.09 & 2.10$\pm$0.10 &8.88 (8.85$-$8.92) & 0.81 (0.74$-$0.87) \\
  49333 &  32504 & 204$\pm$30  &           $-$0.51$\pm$0.32 &  4.216$\pm$0.013 & 2.69$\pm$0.13 & 4.71$\pm$0.23 &7.27 (6.07$-$7.67) & 0.16 (0.00$-$0.43) \\
  49976 &  32838 & 101$\pm$8~~ & \phantom{$-$}1.20$\pm$0.18 &  3.984$\pm$0.014 & 1.49$\pm$0.07 & 2.24$\pm$0.07 &8.45 (8.14$-$8.59) & 0.35 (0.16$-$0.51) \\
  54118 &  34105 &  86$\pm$3~~ & \phantom{$-$}0.38$\pm$0.09 &  4.022$\pm$0.017 & 1.89$\pm$0.05 & 2.72$\pm$0.07 &8.45 (8.34$-$8.52) & 0.61 (0.49$-$0.70) \\
  55522 &  34798 & 220$\pm$29  &           $-$0.87$\pm$0.29 &  4.211$\pm$0.013 & 2.82$\pm$0.12 & 4.88$\pm$0.22 &7.67 (7.36$-$7.79) & 0.46 (0.21$-$0.64) \\
  55719 &  34802 & 133$\pm$8~~ & \phantom{$-$}0.37$\pm$0.14 &  3.960$\pm$0.014 & 1.79$\pm$0.06 & 2.49$\pm$0.07 &8.67 (8.63$-$8.70) & 0.80 (0.73$-$0.85) \\
  56350 &  34929 & 161$\pm$12  & \phantom{$-$}0.64$\pm$0.18 &  4.022$\pm$0.017 & 1.79$\pm$0.08 & 2.61$\pm$0.09 &8.40 (8.19$-$8.51) & 0.48 (0.29$-$0.62) \\
  60435 &  36537 & 233$\pm$45  & \phantom{$-$}1.87$\pm$0.43 &  3.910$\pm$0.011 & 1.14$\pm$0.17 & 1.82$\pm$0.12 &8.82 (8.36$-$8.91) & 0.46 (0.12$-$0.65) \\
  62140 &  37934 &  81$\pm$4~~ & \phantom{$-$}1.89$\pm$0.13 &  3.884$\pm$0.011 & 1.13$\pm$0.05 & 1.77$\pm$0.04 &8.99 (8.93$-$9.04) & 0.62 (0.53$-$0.69) \\
  63401 &  37982 & 210$\pm$24  &           $-$0.40$\pm$0.26 &  4.129$\pm$0.013 & 2.45$\pm$0.11 & 3.79$\pm$0.16 &8.03 (7.90$-$8.11) & 0.58 (0.41$-$0.71) \\
  64486 &  39538 & 101$\pm$5~~ & \phantom{$-$}0.32$\pm$0.12 &  3.999$\pm$0.013 & 1.87$\pm$0.05 & 2.65$\pm$0.06 &8.54 (8.49$-$8.59) & 0.70 (0.62$-$0.77) \\
  64740 &  38500 & 220$\pm$25  &           $-$2.15$\pm$0.25 &  4.353$\pm$0.010 & 3.63$\pm$0.10 & 8.30$\pm$0.30 &7.10 (6.79$-$7.24) & 0.40 (0.18$-$0.57) \\
  65339 &  39261 &  98$\pm$7~~ & \phantom{$-$}1.12$\pm$0.17 &  3.919$\pm$0.010 & 1.45$\pm$0.07 & 2.08$\pm$0.06 &8.84 (8.81$-$8.87) & 0.72 (0.64$-$0.78) \\
  71866 &  41782 & 146$\pm$18  & \phantom{$-$}0.83$\pm$0.28 &  3.944$\pm$0.015 & 1.58$\pm$0.11 & 2.24$\pm$0.11 &8.74 (8.68$-$8.79) & 0.71 (0.58$-$0.81) \\
  72968 &  42146 &  82$\pm$5~~ & \phantom{$-$}1.11$\pm$0.15 &  3.992$\pm$0.013 & 1.54$\pm$0.07 & 2.30$\pm$0.06 &8.43 (8.18$-$8.55) & 0.36 (0.19$-$0.50) \\
  73340 &  42177 & 143$\pm$9~~ &           $-$0.10$\pm$0.14 &  4.145$\pm$0.012 & 2.37$\pm$0.06 & 3.77$\pm$0.11 &7.79 (7.45$-$7.95) & 0.32 (0.14$-$0.47) \\
  74521 &  42917 & 125$\pm$13  & \phantom{$-$}0.07$\pm$0.23 &  4.033$\pm$0.016 & 2.04$\pm$0.10 & 2.92$\pm$0.12 &8.42 (8.35$-$8.47) & 0.69 (0.56$-$0.79) \\
  75445 &  43257 & 113$\pm$8~~ & \phantom{$-$}1.79$\pm$0.16 &  3.885$\pm$0.011 & 1.17$\pm$0.06 & 1.80$\pm$0.05 &8.99 (8.93$-$9.03) & 0.65 (0.56$-$0.72) \\
  79158 &  45290 & 175$\pm$25  &           $-$0.94$\pm$0.31 &  4.097$\pm$0.014 & 2.59$\pm$0.13 & 3.91$\pm$0.23 &8.16 (8.12$-$8.20) & 0.84 (0.74$-$0.92) \\
  81009 &  45999 & 138$\pm$15  & \phantom{$-$}1.17$\pm$0.24 &  3.900$\pm$0.011 & 1.42$\pm$0.10 & 2.04$\pm$0.09 &8.90 (8.86$-$8.94) & 0.78 (0.69$-$0.84) \\
  83368 &  47145 &  72$\pm$3~~ & \phantom{$-$}1.91$\pm$0.12 &  3.877$\pm$0.012 & 1.12$\pm$0.05 & 1.76$\pm$0.04 &9.02 (8.96$-$9.07) & 0.65 (0.56$-$0.72) \\
  83625 &  47272 & 191$\pm$23  & \phantom{$-$}0.35$\pm$0.26 &  4.082$\pm$0.014 & 2.04$\pm$0.11 & 3.08$\pm$0.12 &8.07 (7.63$-$8.24) & 0.36 (0.12$-$0.56) \\
  86199 &  48643 & 235$\pm$28  &           $-$0.27$\pm$0.27 &  4.112$\pm$0.013 & 2.36$\pm$0.11 & 3.58$\pm$0.15 &8.10 (7.95$-$8.16) & 0.58 (0.40$-$0.71) \\
  88158 &  49642 & 250$\pm$32  &           $-$0.72$\pm$0.28 &  4.113$\pm$0.013 & 2.54$\pm$0.12 & 3.87$\pm$0.20 &8.12 (8.08$-$8.16) & 0.75 (0.63$-$0.84) \\
  88385 &  49791 & 273$\pm$47  & \phantom{$-$}0.66$\pm$0.38 &  4.030$\pm$0.016 & 1.80$\pm$0.16 & 2.64$\pm$0.15 &8.34 (7.88$-$8.47) & 0.43 (0.13$-$0.64) \\
  89103 &  50248 & 203$\pm$27  & \phantom{$-$}1.19$\pm$0.29 &  4.069$\pm$0.015 & 1.67$\pm$0.12 & 2.68$\pm$0.11 &6.68 (6.59$-$7.64) & 0.00 (0.00$-$0.08) \\
  90044 &  50885 & 107$\pm$8~~ & \phantom{$-$}0.74$\pm$0.17 &  4.002$\pm$0.013 & 1.71$\pm$0.07 & 2.48$\pm$0.07 &8.49 (8.36$-$8.56) & 0.52 (0.38$-$0.63) \\
  90569 &  51213 & 118$\pm$11  & \phantom{$-$}0.64$\pm$0.21 &  4.003$\pm$0.013 & 1.75$\pm$0.09 & 2.52$\pm$0.09 &8.50 (8.39$-$8.57) & 0.56 (0.42$-$0.68) \\
  92385 &  52059 & 147$\pm$12  & \phantom{$-$}0.81$\pm$0.18 &  4.043$\pm$0.016 & 1.77$\pm$0.08 & 2.66$\pm$0.09 &8.12 (7.28$-$8.35) & 0.26 (0.03$-$0.46) \\
  92499 &  52218 & 224$\pm$44  & \phantom{$-$}2.10$\pm$0.44 &  3.858$\pm$0.012 & 1.05$\pm$0.17 & 1.68$\pm$0.13 &9.10 (9.02$-$9.14) & 0.68 (0.49$-$0.81) \\
  92664 &  52221 & 142$\pm$10  &           $-$0.32$\pm$0.16 &  4.155$\pm$0.012 & 2.48$\pm$0.07 & 3.97$\pm$0.12 &7.84 (7.62$-$7.96) & 0.41 (0.24$-$0.55) \\
  94427 &  53290 & 110$\pm$11  & \phantom{$-$}2.02$\pm$0.22 &  3.861$\pm$0.012 & 1.08$\pm$0.09 & 1.71$\pm$0.07 &9.09 (9.04$-$9.13) & 0.69 (0.59$-$0.78) \\
  94660 &  53379 & 151$\pm$15  & \phantom{$-$}0.20$\pm$0.22 &  4.032$\pm$0.016 & 1.98$\pm$0.09 & 2.85$\pm$0.11 &8.42 (8.33$-$8.48) & 0.64 (0.50$-$0.75) \\
  96707 &  54540 & 108$\pm$7~~ & \phantom{$-$}0.87$\pm$0.14 &  3.893$\pm$0.011 & 1.54$\pm$0.06 & 2.16$\pm$0.06 &8.87 (8.84$-$8.90) & 0.86 (0.82$-$0.90) \\
  98088 &  55106 & 129$\pm$12  & \phantom{$-$}0.79$\pm$0.21 &  3.899$\pm$0.011 & 1.57$\pm$0.09 & 2.20$\pm$0.09 &8.85 (8.81$-$8.89) & 0.86 (0.81$-$0.91) \\
  98340 &  55181 & 226$\pm$37  & \phantom{$-$}0.86$\pm$0.35 &  4.028$\pm$0.016 & 1.71$\pm$0.15 & 2.56$\pm$0.14 &8.24 (6.82$-$8.45) & 0.32 (0.00$-$0.56) \\
 101065 &  56709 & 125$\pm$16  & \phantom{$-$}2.48$\pm$0.29 &  3.810$\pm$0.013 & 0.91$\pm$0.12 & 1.53$\pm$0.09 &9.32 (9.25$-$9.40) & 0.86 (0.76$-$0.94) \\
 103192 &  57936 & 111$\pm$11  &           $-$0.57$\pm$0.21 &  4.044$\pm$0.016 & 2.32$\pm$0.09 & 3.33$\pm$0.14 &8.34 (8.30$-$8.39) & 0.85 (0.77$-$0.91) \\
 105382 &  59173 & 115$\pm$9~~ &           $-$0.93$\pm$0.18 &  4.212$\pm$0.013 & 2.85$\pm$0.08 & 4.93$\pm$0.17 &7.69 (7.49$-$7.78) & 0.49 (0.31$-$0.62) \\
 105770 &  59404 & 194$\pm$23  & \phantom{$-$}0.09$\pm$0.27 &  4.115$\pm$0.013 & 2.22$\pm$0.11 & 3.43$\pm$0.14 &7.94 (7.49$-$8.11) & 0.36 (0.12$-$0.55) \\
 108662 &  60904 &  82$\pm$5~~ & \phantom{$-$}0.57$\pm$0.15 &  4.021$\pm$0.017 & 1.81$\pm$0.07 & 2.63$\pm$0.08 &8.42 (8.26$-$8.52) & 0.52 (0.35$-$0.65) \\
 108945 &  61071 &  95$\pm$7~~ & \phantom{$-$}0.54$\pm$0.17 &  3.952$\pm$0.015 & 1.71$\pm$0.07 & 2.39$\pm$0.08 &8.70 (8.66$-$8.74) & 0.78 (0.69$-$0.84) \\
 109026 &  61199 &  99$\pm$5~~ &           $-$1.28$\pm$0.12 &  4.173$\pm$0.012 & 2.90$\pm$0.05 & 4.79$\pm$0.12 &7.91 (7.87$-$7.95) & 0.78 (0.71$-$0.83) \\
 110066 &  61748 & 155$\pm$17  & \phantom{$-$}0.41$\pm$0.25 &  3.947$\pm$0.015 & 1.75$\pm$0.10 & 2.44$\pm$0.12 &8.71 (8.66$-$8.74) & 0.82 (0.73$-$0.89) \\
 111133 &  62376 & 160$\pm$23  & \phantom{$-$}0.20$\pm$0.33 &  3.997$\pm$0.013 & 1.92$\pm$0.13 & 2.69$\pm$0.16 &8.55 (8.50$-$8.58) & 0.75 (0.62$-$0.84) \\
 112185 &  62956 &  24$\pm$0~~ &           $-$0.21$\pm$0.04 &  3.953$\pm$0.015 & 2.01$\pm$0.02 & 2.76$\pm$0.03 &8.61 (8.59$-$8.63) & 0.94 (0.90$-$0.97) \\
 112381 &  63204 & 100$\pm$8~~ & \phantom{$-$}1.53$\pm$0.19 &  3.999$\pm$0.013 & 1.39$\pm$0.08 & 2.20$\pm$0.07 &7.02 (6.98$-$8.26) & 0.00 (0.00$-$0.22) \\
 112413 &  63125 &  33$\pm$1~~ & \phantom{$-$}0.26$\pm$0.08 &  4.060$\pm$0.015 & 2.03$\pm$0.05 & 2.98$\pm$0.07 &8.27 (8.14$-$8.36) & 0.52 (0.39$-$0.63) \\
 115226 &  64883 & 141$\pm$18  & \phantom{$-$}2.57$\pm$0.29 &  3.883$\pm$0.011 & 0.86$\pm$0.12 & 1.60$\pm$0.05 &8.60 (7.37$-$9.16) & 0.18 (0.00$-$0.62) \\
 115440 &  65053 & 229$\pm$40  & \phantom{$-$}0.81$\pm$0.39 &  4.086$\pm$0.014 & 1.86$\pm$0.16 & 2.91$\pm$0.15 &6.51 (6.45$-$8.04) & 0.00 (0.00$-$0.30) \\
 115708 &  64936 & 132$\pm$18  & \phantom{$-$}2.19$\pm$0.31 &  3.880$\pm$0.011 & 1.01$\pm$0.12 & 1.68$\pm$0.08 &8.97 (8.76$-$9.04) & 0.50 (0.28$-$0.65) \\
 116114 &  65203 & 140$\pm$17  & \phantom{$-$}1.41$\pm$0.28 &  3.870$\pm$0.012 & 1.32$\pm$0.11 & 1.92$\pm$0.10 &9.01 (8.96$-$9.05) & 0.82 (0.74$-$0.89) \\
 116458 &  65522 & 142$\pm$11  &           $-$0.15$\pm$0.18 &  4.012$\pm$0.013 & 2.08$\pm$0.07 & 2.93$\pm$0.10 &8.48 (8.44$-$8.51) & 0.81 (0.74$-$0.87) \\
 116890 &  65755 & 214$\pm$26  &           $-$0.94$\pm$0.27 &  4.112$\pm$0.013 & 2.63$\pm$0.11 & 4.01$\pm$0.20 &8.12 (8.08$-$8.15) & 0.81 (0.72$-$0.89) \\
 117025 &  65783 &  88$\pm$4~~ & \phantom{$-$}1.25$\pm$0.13 &  3.945$\pm$0.015 & 1.42$\pm$0.05 & 2.09$\pm$0.05 &8.72 (8.61$-$8.80) & 0.55 (0.43$-$0.65) \\
 118022 &  66200 &  56$\pm$2~~ & \phantom{$-$}1.17$\pm$0.10 &  3.957$\pm$0.014 & 1.46$\pm$0.04 & 2.16$\pm$0.05 &8.66 (8.56$-$8.74) & 0.52 (0.41$-$0.62) \\
 118913 &  66888 & 206$\pm$31  & \phantom{$-$}0.73$\pm$0.33 &  3.981$\pm$0.014 & 1.67$\pm$0.14 & 2.40$\pm$0.14 &8.60 (8.48$-$8.66) & 0.61 (0.43$-$0.74) \\
 119213 &  66700 &  88$\pm$4~~ & \phantom{$-$}1.52$\pm$0.12 &  3.941$\pm$0.015 & 1.30$\pm$0.05 & 1.99$\pm$0.05 &8.69 (8.51$-$8.80) & 0.43 (0.29$-$0.55) \\
 119308 &  66942 & 182$\pm$31  & \phantom{$-$}1.35$\pm$0.37 &  4.009$\pm$0.013 & 1.48$\pm$0.15 & 2.30$\pm$0.11 &7.73 (6.93$-$8.40) & 0.06 (0.00$-$0.36) \\
 119419 &  67036 & 112$\pm$9~~ & \phantom{$-$}1.16$\pm$0.18 &  4.048$\pm$0.016 & 1.64$\pm$0.08 & 2.55$\pm$0.09 &6.78 (6.71$-$8.05) & 0.00 (0.00$-$0.20) \\
 120198 &  67231 &  86$\pm$4~~ & \phantom{$-$}0.98$\pm$0.12 &  4.023$\pm$0.016 & 1.65$\pm$0.06 & 2.49$\pm$0.07 &8.20 (7.59$-$8.41) & 0.26 (0.06$-$0.43) \\
 122532 &  68673 & 169$\pm$21  &           $-$0.15$\pm$0.28 &  4.071$\pm$0.015 & 2.22$\pm$0.12 & 3.25$\pm$0.15 &8.27 (8.18$-$8.33) & 0.67 (0.51$-$0.78) \\
 122970 &  68790 & 129$\pm$16  & \phantom{$-$}2.74$\pm$0.28 &  3.840$\pm$0.013 & 0.79$\pm$0.11 & 1.48$\pm$0.07 &9.20 (8.99$-$9.29) & 0.57 (0.33$-$0.71) \\
 124224 &  69389 &  80$\pm$5~~ & \phantom{$-$}0.46$\pm$0.16 &  4.084$\pm$0.014 & 2.00$\pm$0.07 & 3.04$\pm$0.09 &7.97 (7.45$-$8.18) & 0.27 (0.07$-$0.45) \\
 125248 &  69929 &  90$\pm$7~~ & \phantom{$-$}1.20$\pm$0.18 &  3.992$\pm$0.013 & 1.50$\pm$0.08 & 2.27$\pm$0.07 &8.37 (7.94$-$8.53) & 0.30 (0.10$-$0.47) \\
 125630 &  70346 & 159$\pm$16  & \phantom{$-$}0.55$\pm$0.23 &  3.966$\pm$0.014 & 1.72$\pm$0.10 & 2.42$\pm$0.10 &8.66 (8.61$-$8.70) & 0.73 (0.62$-$0.81) \\
 125823 &  70300 & 128$\pm$12  &           $-$1.18$\pm$0.21 &  4.248$\pm$0.012 & 3.02$\pm$0.09 & 5.57$\pm$0.20 &7.51 (7.25$-$7.63) & 0.45 (0.24$-$0.59) \\
 126515 &  70553 & 141$\pm$21  & \phantom{$-$}1.26$\pm$0.32 &  4.007$\pm$0.013 & 1.51$\pm$0.13 & 2.32$\pm$0.11 &8.07 (6.93$-$8.44) & 0.16 (0.00$-$0.41) \\
 127453 &  71314 & 257$\pm$49  &           $-$0.17$\pm$0.42 &  4.083$\pm$0.014 & 2.25$\pm$0.17 & 3.33$\pm$0.23 &8.22 (8.07$-$8.28) & 0.64 (0.41$-$0.79) \\
 127575 &  71359 & 183$\pm$28  & \phantom{$-$}0.91$\pm$0.34 &  4.081$\pm$0.014 & 1.81$\pm$0.14 & 2.84$\pm$0.13 &6.55 (6.49$-$7.98) & 0.00 (0.00$-$0.24) \\
 128775 &  71727 & 161$\pm$22  & \phantom{$-$}0.48$\pm$0.30 &  4.076$\pm$0.015 & 1.97$\pm$0.12 & 2.98$\pm$0.13 &8.04 (7.12$-$8.25) & 0.30 (0.03$-$0.53) \\
 128898 &  71908 &  16$\pm$0~~ & \phantom{$-$}2.12$\pm$0.03 &  3.885$\pm$0.011 & 1.04$\pm$0.01 & 1.71$\pm$0.02 &8.95 (8.86$-$9.03) & 0.51 (0.42$-$0.58) \\
 129899 &  72670 & 275$\pm$43  &           $-$0.94$\pm$0.34 &  4.017$\pm$0.013 & 2.41$\pm$0.14 & 3.43$\pm$0.19 &8.36 (8.31$-$8.41) & 0.95 (0.89$-$1.00) \\
 130559 &  72489 &  72$\pm$6~~ & \phantom{$-$}1.32$\pm$0.20 &  3.957$\pm$0.014 & 1.40$\pm$0.08 & 2.10$\pm$0.07 &8.62 (8.42$-$8.72) & 0.44 (0.26$-$0.58) \\
 132322 &  73520 & 173$\pm$21  & \phantom{$-$}0.93$\pm$0.28 &  3.922$\pm$0.010 & 1.52$\pm$0.11 & 2.16$\pm$0.11 &8.82 (8.78$-$8.86) & 0.76 (0.67$-$0.83) \\
 133029 &  73454 & 146$\pm$11  & \phantom{$-$}0.48$\pm$0.18 &  4.027$\pm$0.016 & 1.86$\pm$0.08 & 2.70$\pm$0.09 &8.41 (8.26$-$8.49) & 0.54 (0.37$-$0.67) \\
 133652 &  73937 &  95$\pm$8~~ & \phantom{$-$}0.81$\pm$0.20 &  4.113$\pm$0.013 & 1.93$\pm$0.09 & 3.10$\pm$0.10 &6.39 (6.34$-$7.21) & 0.00 (0.00$-$0.04) \\
 133792 &  74181 & 170$\pm$19  &           $-$0.17$\pm$0.24 &  3.974$\pm$0.014 & 2.02$\pm$0.10 & 2.80$\pm$0.14 &8.58 (8.54$-$8.62) & 0.89 (0.82$-$0.94) \\
 133880 &  74066 & 126$\pm$13  & \phantom{$-$}0.25$\pm$0.23 &  4.079$\pm$0.014 & 2.07$\pm$0.10 & 3.10$\pm$0.12 &8.15 (7.88$-$8.27) & 0.44 (0.22$-$0.60) \\
 134214 &  74145 &  91$\pm$7~~ & \phantom{$-$}2.58$\pm$0.18 &  3.858$\pm$0.012 & 0.85$\pm$0.07 & 1.55$\pm$0.04 &9.05 (8.84$-$9.16) & 0.47 (0.28$-$0.61) \\
 134305 &  74109 & 177$\pm$27  & \phantom{$-$}0.93$\pm$0.33 &  3.908$\pm$0.011 & 1.52$\pm$0.13 & 2.15$\pm$0.14 &8.85 (8.80$-$8.90) & 0.81 (0.72$-$0.88) \\
 137509 &  76011 & 249$\pm$37  &           $-$0.28$\pm$0.33 &  4.147$\pm$0.012 & 2.44$\pm$0.14 & 3.88$\pm$0.18 &7.89 (7.51$-$8.02) & 0.43 (0.17$-$0.62) \\
 137909 &  75695 &  34$\pm$0~~ & \phantom{$-$}1.12$\pm$0.06 &  3.871$\pm$0.012 & 1.44$\pm$0.02 & 2.04$\pm$0.02 &8.95 (8.93$-$8.97) & 0.88 (0.84$-$0.91) \\
 137949 &  75848 &  89$\pm$6~~ & \phantom{$-$}1.80$\pm$0.17 &  3.861$\pm$0.012 & 1.16$\pm$0.07 & 1.78$\pm$0.06 &9.07 (9.03$-$9.11) & 0.75 (0.68$-$0.82) \\
 138758 &  76767 & 238$\pm$44  & \phantom{$-$}0.92$\pm$0.41 &  4.023$\pm$0.016 & 1.68$\pm$0.17 & 2.51$\pm$0.15 &8.26 (6.86$-$8.47) & 0.31 (0.00$-$0.57) \\
 140160 &  76866 &  70$\pm$4~~ & \phantom{$-$}1.04$\pm$0.13 &  3.968$\pm$0.014 & 1.53$\pm$0.06 & 2.24$\pm$0.06 &8.62 (8.52$-$8.70) & 0.53 (0.41$-$0.63) \\
 140728 &  76957 &  97$\pm$4~~ & \phantom{$-$}0.51$\pm$0.11 &  4.021$\pm$0.012 & 1.84$\pm$0.05 & 2.66$\pm$0.06 &8.43 (8.35$-$8.50) & 0.55 (0.44$-$0.64) \\
 142301 &  77909 & 139$\pm$23  &           $-$0.24$\pm$0.37 &  4.193$\pm$0.011 & 2.53$\pm$0.15 & 4.30$\pm$0.21 &7.21 (6.14$-$7.72) & 0.11 (0.00$-$0.40) \\
 142990 &  78246 & 149$\pm$18  &           $-$0.74$\pm$0.27 &  4.217$\pm$0.013 & 2.78$\pm$0.11 & 4.87$\pm$0.20 &7.54 (6.98$-$7.73) & 0.34 (0.08$-$0.54) \\
 143473 &  78533 & 123$\pm$15  & \phantom{$-$}1.06$\pm$0.27 &  4.109$\pm$0.014 & 1.82$\pm$0.11 & 3.00$\pm$0.12 &6.44 (6.38$-$6.52) & 0.00 (0.00$-$0.00) \\
 144334 &  78877 & 149$\pm$19  &           $-$0.29$\pm$0.28 &  4.168$\pm$0.012 & 2.50$\pm$0.12 & 4.08$\pm$0.17 &7.72 (7.10$-$7.90) & 0.33 (0.07$-$0.52) \\
 145501 &  79374 & 133$\pm$19  & \phantom{$-$}0.03$\pm$0.33 &  4.141$\pm$0.013 & 2.31$\pm$0.13 & 3.67$\pm$0.17 &7.72 (6.26$-$7.98) & 0.25 (0.00$-$0.50) \\
 147010 &  80024 & 143$\pm$19  & \phantom{$-$}0.54$\pm$0.30 &  4.117$\pm$0.013 & 2.04$\pm$0.12 & 3.23$\pm$0.14 &6.47 (6.30$-$7.86) & 0.00 (0.00$-$0.27) \\
 148112 &  80463 &  72$\pm$4~~ & \phantom{$-$}0.20$\pm$0.15 &  3.970$\pm$0.014 & 1.87$\pm$0.06 & 2.60$\pm$0.08 &8.63 (8.59$-$8.66) & 0.81 (0.75$-$0.87) \\
 148199 &  80607 & 150$\pm$21  & \phantom{$-$}0.49$\pm$0.31 &  4.046$\pm$0.016 & 1.90$\pm$0.13 & 2.79$\pm$0.13 &8.29 (7.95$-$8.40) & 0.45 (0.20$-$0.64) \\
 148330 &  80375 & 112$\pm$6~~ & \phantom{$-$}0.49$\pm$0.12 &  3.947$\pm$0.015 & 1.72$\pm$0.05 & 2.40$\pm$0.06 &8.71 (8.67$-$8.74) & 0.80 (0.74$-$0.85) \\
 149764 &  81477 & 126$\pm$15  & \phantom{$-$}0.99$\pm$0.28 &  4.128$\pm$0.013 & 1.89$\pm$0.12 & 3.19$\pm$0.12 &6.35 (6.31$-$6.41) & 0.00 (0.00$-$0.00) \\
 149822 &  81337 & 133$\pm$14  & \phantom{$-$}0.70$\pm$0.23 &  4.010$\pm$0.013 & 1.74$\pm$0.10 & 2.53$\pm$0.09 &8.46 (8.29$-$8.53) & 0.51 (0.33$-$0.64) \\
 149911 &  81440 & 126$\pm$16  & \phantom{$-$}0.06$\pm$0.29 &  3.970$\pm$0.014 & 1.93$\pm$0.12 & 2.67$\pm$0.15 &8.61 (8.57$-$8.65) & 0.85 (0.76$-$0.92) \\
 151525 &  82216 & 141$\pm$17  &           $-$0.64$\pm$0.27 &  3.971$\pm$0.014 & 2.21$\pm$0.11 & 3.04$\pm$0.13 &8.51 (8.46$-$8.55) & 0.97 (0.92$-$1.00) \\
 151965 &  82554 & 180$\pm$28  &           $-$0.09$\pm$0.35 &  4.154$\pm$0.012 & 2.38$\pm$0.14 & 3.85$\pm$0.19 &7.68 (6.25$-$7.94) & 0.26 (0.00$-$0.50) \\
 152107 &  82321 &  53$\pm$1~~ & \phantom{$-$}1.21$\pm$0.07 &  3.941$\pm$0.015 & 1.43$\pm$0.03 & 2.10$\pm$0.04 &8.74 (8.65$-$8.82) & 0.58 (0.48$-$0.67) \\
 153882 &  83308 & 168$\pm$20  &           $-$0.07$\pm$0.27 &  3.988$\pm$0.013 & 2.00$\pm$0.11 & 2.79$\pm$0.14 &8.55 (8.51$-$8.59) & 0.84 (0.76$-$0.91) \\
 154708 &  84017 & 140$\pm$22  & \phantom{$-$}2.90$\pm$0.35 &  3.829$\pm$0.013 & 0.73$\pm$0.14 & 1.43$\pm$0.08 &9.29 (9.06$-$9.38) & 0.61 (0.33$-$0.76) \\
 157751 &  85372 & 164$\pm$28  & \phantom{$-$}1.52$\pm$0.38 &  3.993$\pm$0.013 & 1.38$\pm$0.15 & 2.18$\pm$0.10 &7.56 (7.00$-$8.45) & 0.03 (0.00$-$0.35) \\
 164258 &  88148 & 121$\pm$13  & \phantom{$-$}0.63$\pm$0.24 &  3.952$\pm$0.015 & 1.67$\pm$0.10 & 2.35$\pm$0.11 &8.71 (8.66$-$8.75) & 0.75 (0.64$-$0.83) \\
 165474 &  88627 & 130$\pm$20  & \phantom{$-$}1.75$\pm$0.35 &  3.885$\pm$0.011 & 1.19$\pm$0.14 & 1.82$\pm$0.11 &8.99 (8.92$-$9.03) & 0.66 (0.51$-$0.77) \\
 168733 &  90074 & 189$\pm$27  &           $-$1.14$\pm$0.31 &  4.108$\pm$0.014 & 2.70$\pm$0.13 & 4.13$\pm$0.25 &8.12 (8.08$-$8.16) & 0.87 (0.78$-$0.94) \\
 168856 &  90030 & 179$\pm$29  & \phantom{$-$}0.28$\pm$0.37 &  4.106$\pm$0.014 & 2.13$\pm$0.15 & 3.28$\pm$0.17 &7.89 (6.34$-$8.14) & 0.28 (0.00$-$0.54) \\
 170000 &  89908 &  88$\pm$3~~ &           $-$0.39$\pm$0.10 &  4.058$\pm$0.015 & 2.28$\pm$0.05 & 3.30$\pm$0.08 &8.32 (8.27$-$8.36) & 0.78 (0.70$-$0.84) \\
 170397 &  90651 &  87$\pm$6~~ & \phantom{$-$}1.14$\pm$0.16 &  4.033$\pm$0.016 & 1.61$\pm$0.07 & 2.48$\pm$0.08 &7.70 (6.79$-$8.24) & 0.07 (0.00$-$0.29) \\
 171184 &  91001 & 186$\pm$35  & \phantom{$-$}0.44$\pm$0.41 &  4.081$\pm$0.014 & 2.00$\pm$0.17 & 3.03$\pm$0.18 &8.02 (6.48$-$8.25) & 0.30 (0.00$-$0.56) \\
 171586 &  91142 & 102$\pm$9~~ & \phantom{$-$}1.15$\pm$0.21 &  3.967$\pm$0.014 & 1.48$\pm$0.09 & 2.19$\pm$0.08 &8.60 (8.43$-$8.69) & 0.48 (0.30$-$0.61) \\
 172690 &  93481 & 249$\pm$42  & \phantom{$-$}0.19$\pm$0.37 &  4.058$\pm$0.015 & 2.05$\pm$0.15 & 3.00$\pm$0.18 &8.30 (8.08$-$8.38) & 0.56 (0.32$-$0.72) \\
 173650 &  92036 & 214$\pm$34  &           $-$0.42$\pm$0.35 &  4.015$\pm$0.013 & 2.20$\pm$0.14 & 3.10$\pm$0.20 &8.44 (8.39$-$8.49) & 0.86 (0.78$-$0.93) \\
 175132 &  92599 & 374$\pm$72  &           $-$1.72$\pm$0.43 &  4.022$\pm$0.017 & 2.73$\pm$0.17 & 3.73$\pm$0.18 &8.27 (8.23$-$8.32) & 1.00 (1.00$-$1.00) \\
 175362 &  92989 & 130$\pm$15  &           $-$0.40$\pm$0.27 &  4.217$\pm$0.013 & 2.64$\pm$0.11 & 4.64$\pm$0.20 &6.92 (6.04$-$7.56) & 0.06 (0.00$-$0.32) \\
 176196 &  93863 & 255$\pm$44  & \phantom{$-$}0.47$\pm$0.38 &  4.000$\pm$0.013 & 1.81$\pm$0.15 & 2.58$\pm$0.17 &8.53 (8.43$-$8.59) & 0.64 (0.46$-$0.77) \\
 176232 &  93179 &  74$\pm$3~~ & \phantom{$-$}1.42$\pm$0.11 &  3.899$\pm$0.011 & 1.32$\pm$0.04 & 1.95$\pm$0.04 &8.93 (8.88$-$8.96) & 0.71 (0.64$-$0.77) \\
 179527 &  94311 & 300$\pm$49  &           $-$1.64$\pm$0.36 &  4.046$\pm$0.016 & 2.75$\pm$0.15 & 3.94$\pm$0.18 &8.22 (8.18$-$8.26) & 1.00 (0.96$-$1.00) \\
 183056 &  95556 & 200$\pm$20  &           $-$1.28$\pm$0.22 &  4.086$\pm$0.014 & 2.70$\pm$0.10 & 4.08$\pm$0.18 &8.16 (8.12$-$8.19) & 0.93 (0.87$-$0.97) \\
 183339 &  95520 & 384$\pm$76  &           $-$1.35$\pm$0.43 &  4.126$\pm$0.013 & 2.82$\pm$0.18 & 4.44$\pm$0.37 &8.05 (7.99$-$8.10) & 0.88 (0.78$-$0.96) \\
 183806 &  96178 & 133$\pm$15  &           $-$0.09$\pm$0.25 &  3.983$\pm$0.014 & 2.00$\pm$0.10 & 2.79$\pm$0.14 &8.56 (8.52$-$8.60) & 0.85 (0.78$-$0.92) \\
 184905 &  96292 & 165$\pm$13  & \phantom{$-$}0.34$\pm$0.18 &  4.035$\pm$0.016 & 1.93$\pm$0.08 & 2.80$\pm$0.09 &8.39 (8.26$-$8.47) & 0.57 (0.41$-$0.69) \\
 187474 &  97749 & 103$\pm$9~~ & \phantom{$-$}0.32$\pm$0.20 &  4.004$\pm$0.013 & 1.88$\pm$0.08 & 2.67$\pm$0.09 &8.52 (8.47$-$8.57) & 0.69 (0.58$-$0.77) \\
 188041 &  97871 &  84$\pm$6~~ & \phantom{$-$}0.86$\pm$0.17 &  3.926$\pm$0.010 & 1.55$\pm$0.07 & 2.20$\pm$0.07 &8.80 (8.77$-$8.83) & 0.77 (0.70$-$0.82) \\
 192678 &  99672 & 230$\pm$27  & \phantom{$-$}0.41$\pm$0.26 &  3.987$\pm$0.013 & 1.81$\pm$0.10 & 2.56$\pm$0.12 &8.59 (8.54$-$8.63) & 0.71 (0.59$-$0.80) \\
 196178 & 101475 & 147$\pm$14  &           $-$0.16$\pm$0.22 &  4.126$\pm$0.013 & 2.35$\pm$0.09 & 3.64$\pm$0.13 &7.99 (7.76$-$8.09) & 0.47 (0.27$-$0.61) \\
 196502 & 101260 & 127$\pm$8~~ &           $-$0.36$\pm$0.15 &  3.964$\pm$0.014 & 2.08$\pm$0.07 & 2.87$\pm$0.09 &8.57 (8.54$-$8.60) & 0.94 (0.90$-$1.00) \\
 199728 & 103616 & 131$\pm$17  & \phantom{$-$}0.60$\pm$0.30 &  4.078$\pm$0.015 & 1.93$\pm$0.12 & 2.95$\pm$0.13 &7.89 (6.51$-$8.21) & 0.21 (0.00$-$0.45) \\
 200177 & 103658 & 139$\pm$11  & \phantom{$-$}1.47$\pm$0.18 &  3.997$\pm$0.013 & 1.41$\pm$0.08 & 2.21$\pm$0.07 &7.70 (6.99$-$8.34) & 0.05 (0.00$-$0.27) \\
 201018 & 104337 & 174$\pm$32  & \phantom{$-$}2.32$\pm$0.40 &  3.941$\pm$0.015 & 0.98$\pm$0.16 & 1.82$\pm$0.07 &7.05 (7.05$-$8.41) & 0.00 (0.00$-$0.17) \\
 201601 & 104521 &  35$\pm$1~~ & \phantom{$-$}1.98$\pm$0.07 &  3.882$\pm$0.011 & 1.10$\pm$0.03 & 1.74$\pm$0.03 &8.99 (8.93$-$9.05) & 0.59 (0.52$-$0.66) \\
 203006 & 105382 &  57$\pm$2~~ & \phantom{$-$}1.13$\pm$0.11 &  3.989$\pm$0.013 & 1.52$\pm$0.05 & 2.28$\pm$0.05 &8.44 (8.23$-$8.56) & 0.37 (0.22$-$0.49) \\
 204411 & 105898 & 119$\pm$7~~ &           $-$0.10$\pm$0.14 &  3.942$\pm$0.015 & 1.95$\pm$0.06 & 2.68$\pm$0.07 &8.65 (8.62$-$8.68) & 0.94 (0.89$-$0.98) \\
 205087 & 106355 & 186$\pm$23  & \phantom{$-$}0.20$\pm$0.28 &  4.038$\pm$0.016 & 2.00$\pm$0.12 & 2.88$\pm$0.14 &8.39 (8.28$-$8.46) & 0.63 (0.45$-$0.75) \\
 208217 & 108340 & 146$\pm$19  & \phantom{$-$}1.50$\pm$0.29 &  3.904$\pm$0.011 & 1.29$\pm$0.12 & 1.93$\pm$0.10 &8.91 (8.85$-$8.95) & 0.66 (0.52$-$0.76) \\
 212385 & 110624 & 112$\pm$11  & \phantom{$-$}1.58$\pm$0.23 &  3.923$\pm$0.010 & 1.26$\pm$0.09 & 1.93$\pm$0.07 &8.80 (8.69$-$8.86) & 0.51 (0.37$-$0.63) \\
 215038 & 111849 & 265$\pm$42  & \phantom{$-$}0.57$\pm$0.35 &  4.136$\pm$0.013 & 2.08$\pm$0.14 & 3.39$\pm$0.14 &6.29 (6.26$-$7.60) & 0.00 (0.00$-$0.16) \\
 216018 & 112705 & 109$\pm$11  & \phantom{$-$}2.51$\pm$0.24 &  3.889$\pm$0.011 & 0.88$\pm$0.10 & 1.63$\pm$0.04 &8.54 (7.29$-$8.90) & 0.16 (0.00$-$0.40) \\
 217522 & 113711 &  95$\pm$8~~ & \phantom{$-$}2.62$\pm$0.20 &  3.816$\pm$0.013 & 0.85$\pm$0.08 & 1.49$\pm$0.06 &9.33 (9.27$-$9.40) & 0.80 (0.69$-$0.87) \\
 217833 & 113797 & 221$\pm$40  &           $-$0.52$\pm$0.40 &  4.171$\pm$0.012 & 2.59$\pm$0.16 & 4.23$\pm$0.24 &7.82 (7.41$-$7.94) & 0.45 (0.16$-$0.66) \\
 220825 & 115738 &  49$\pm$1~~ & \phantom{$-$}1.45$\pm$0.09 &  3.958$\pm$0.014 & 1.35$\pm$0.04 & 2.07$\pm$0.04 &8.57 (8.36$-$8.70) & 0.37 (0.23$-$0.48) \\
 221006 & 115908 & 116$\pm$7~~ & \phantom{$-$}0.29$\pm$0.14 &  4.135$\pm$0.013 & 2.19$\pm$0.06 & 3.49$\pm$0.10 &7.32 (6.26$-$7.80) & 0.08 (0.00$-$0.27) \\
 221394 & 116119 & 147$\pm$14  & \phantom{$-$}0.45$\pm$0.22 &  3.979$\pm$0.014 & 1.78$\pm$0.09 & 2.51$\pm$0.10 &8.62 (8.56$-$8.66) & 0.72 (0.61$-$0.80) \\
 221568 & 116210 & 243$\pm$39  & \phantom{$-$}0.24$\pm$0.36 &  3.962$\pm$0.014 & 1.84$\pm$0.14 & 2.56$\pm$0.17 &8.65 (8.60$-$8.70) & 0.82 (0.72$-$0.91) \\
 223640 & 117629 &  98$\pm$10  & \phantom{$-$}0.17$\pm$0.22 &  4.089$\pm$0.014 & 2.13$\pm$0.10 & 3.20$\pm$0.12 &8.11 (7.84$-$8.23) & 0.44 (0.23$-$0.60) \\
 224801 &     63 & 207$\pm$30  &           $-$0.43$\pm$0.32 &  4.073$\pm$0.015 & 2.33$\pm$0.13 & 3.41$\pm$0.20 &8.27 (8.21$-$8.31) & 0.76 (0.62$-$0.85) \\
\end{longtable}
}

\end{document}